\documentclass[journal,draftclsnofoot,onecolumn,12pt]{IEEEtran}

\usepackage{amsthm,amssymb,graphicx,multirow,amsmath,color,amsfonts,mathtools}
\usepackage[update,prepend]{epstopdf}
\usepackage[noadjust]{cite}
\usepackage[latin1]{inputenc}
\usepackage{tikz}
\usetikzlibrary{arrows,calc} 
\usepackage{bbm} 
\usepackage{pdfpages}
\usepackage{tabulary}
\usepackage{multirow}
\usepackage{comment}
\usepackage{textcomp}

\def\nbo{{\mathbf{o}}}

\def\nbx{{\mathbf{x}}}

\def\nb0{{\mathbf{0}}}
\def\nb1{{\mathbf{1}}}

\def\ncalA{{\mathcal{A}}}
\def\ncalB{{\mathcal{B}}}

\def\ncalL{{\mathcal{L}}}

\def\ncalR{{\mathcal{R}}}
\def\ncalS{{\mathcal{S}}}

\def\ncalX{{\mathcal{X}}}

\def\nbbE{{\mathbb{E}}}

\def\nbbP{{\mathbb{P}}}

\newtheorem{lemma}{Lemma}

\newtheorem{theorem}{Theorem}

\newtheorem{cor}{Corollary}

\newtheorem{remark}{Remark}
\newtheorem{assumption}{Assumption}
	
\def\argmin{\operatorname{arg~min}}
\def\argmax{\operatorname{arg~max}}
\allowdisplaybreaks 

\begin{document}
\graphicspath{{./Figures/}}
\title{3D Two-Hop Cellular Networks with Wireless Backhauled UAVs: Modeling and Fundamentals}
\author{
Morteza Banagar and Harpreet S. Dhillon
\thanks{The authors are with Wireless@VT, Department of ECE, Virginia Tech, Blacksburg, VA. Email: \{mbanagar, hdhillon\}@vt.edu. The support of the US NSF (Grants CNS-1923807 and AST-2037870) is gratefully acknowledged. A part of this work is submitted to the 2021 IEEE Globecom in Madrid, Spain \cite{C_Morteza_Fundamentals_2021}.\hfill Manuscript updated: \today.}}

\maketitle
\vspace{-1.4cm}
\begin{abstract}
In this paper, we characterize the performance of a three-dimensional (3D) two-hop cellular network in which terrestrial base stations (BSs) coexist with unmanned aerial vehicles (UAVs) to serve a set of ground user equipment (UE). In particular, a UE connects either directly to its serving terrestrial BS by an access link or connects first to its serving UAV which is then wirelessly backhauled to a terrestrial BS (joint access and backhaul). We consider realistic antenna radiation patterns for both BSs and UAVs using practical models developed by the third generation partnership project (3GPP). We assume a probabilistic channel model for the air-to-ground transmission, which incorporates both line-of-sight (LoS) and non-line-of-sight (NLoS) links. Assuming the max-power association policy, we study the performance of the network in both amplify-and-forward (AF) and decode-and-forward (DF) relaying protocols. Using tools from stochastic geometry, we analyze the joint distribution of distance and zenith angle of the closest (and serving) UAV to the origin in a 3D setting. Further, we identify and extensively study key mathematical constructs as the building blocks of characterizing the received signal-to-interference-plus-noise ratio (SINR) distribution. Using these results, we obtain exact mathematical expressions for the coverage probability in both AF and DF relaying protocols. Furthermore, considering the fact that backhaul links could be quite weak because of the downtilted antennas at the BSs, we propose and analyze the addition of a directional uptilted antenna at the BS that is solely used for backhaul purposes. The superiority of having directional antennas with wirelessly backhauled UAVs is further demonstrated via simulation.
\end{abstract}

\vspace{-0.4cm}
\begin{IEEEkeywords}
\vspace{-0.4cm}
Unmanned aerial vehicle, wireless backhaul, stochastic geometry, amplify-and-forward, decode-and-forward, aerial-terrestrial coexistence.
\end{IEEEkeywords}

\vspace{-0.4cm}
\section{Introduction} \label{sec:intro}
Owing to their unique deployment advantages, such as agility, cost-effectiveness, and high probability of LoS, UAVs are widely regarded as an indispensable component of modern wireless networks \cite{J_Zeng_Wireless_2016, J_Mozaffari_Tutorial_2018, J_Morteza_Performance_2019}. Specifically, UAVs can act as aerial UEs, BSs, or even relays to expand the coverage or capacity of a terrestrial network or to establish a temporary wireless network in case of natural disasters. Since most UAV platforms are able to move freely in the sky and do not have any wired connection to the ground, they will naturally need to establish \textit{wireless backhaul} connections with fiber-backhauled BSs on the ground. Since terrestrial BSs are downtilted, these wireless backhaul connections may be established through BS antenna sidelobes, which may limit their capacity \cite{C_Fakhreddine_Handover_2019, C_Amer_Performance_2020}. Given the 3D nature of this network and the intricate dependencies of access and backhaul links in this setup, it is not straightforward to quantify the performance of this network. Tackling this important challenge, this paper develops a comprehensive framework with foundations in stochastic geometry to study the end-to-end performance of this 3D two-hop network in which the UE on the ground may be served by a UAV, which, in turn, is wirelessly backhauled to a terrestrial BS. Crucially, our analysis is cognizant of the performance limiting characteristics of the UAV networks, such as the realistic antenna patterns of BSs. Going further, we also characterize the performance gains obtained by deploying dedicated uptilted antennas at the terrestrial BSs specifically for the backhaul purposes. 


\vspace{-0.4cm}
\subsection{Related Works} \label{sec:related}
This paper lies at the intersection of the following three research directions: (i) relaying in cellular networks, (ii) stochastic geometry for UAV networks, and (iii) wirelessly backhauled UAVs. We discuss each of the above lines of research next.

{\em Relaying in Cellular Networks.} The idea of using relays for improving the performance of cellular networks, such as increasing the coverage area or offering higher throughput to the UEs, is well-established in wireless communications \cite{J_Nosratinia_Cooperative_2004, J_Laneman_Cooperative_2004, J_Hasna_Outage_2003, J_Hasna_End_2003, J_Cho_Throughput_2004, J_Senaratne_Unified_2010}. Two major cooperative signaling methods, i.e., AF and DF, have been extensively studied in the literature \cite{J_Nosratinia_Cooperative_2004}. For instance, the authors in \cite{J_Laneman_Cooperative_2004} studied the information-theoretic aspects of both AF and DF relaying schemes along with their variants. End-to-end performance of cooperative relay networks over Rayleigh and Nakagami fading channels has been analyzed in several works, such as in \cite{J_Hasna_Outage_2003, J_Hasna_End_2003, J_Senaratne_Unified_2010}, where the authors obtained mathematical expressions for the end-to-end signal-to-noise ratio (SNR) of each relaying protocol. Moving forward to the past decade, interference has become a non-negligible factor in determining the performance of wireless networks due to a dramatic increase in the number of nodes and bandwidth scarcity. Taking the impact of interference into account, the authors in \cite{J_Aalo_Performance_2014} studied the outage performance of a multi-hop AF communication system where the relays were exposed to a Poisson field of interferes. Using tools from stochastic geometry and optimization theory, the authors in \cite{J_Lu_Stochastic_2015} provided a system-level analysis of two-hop DF networks and showed that the benefits of relays could be negligible if the system is not appropriately designed. In recent years and with the emergence of UAVs as potential wireless nodes, there has been a lot of interest in using UAVs as relays \cite{J_Zhang_Joint_2018, J_Chen_Multiple_2018, J_Chen_Optimum_2018, J_Pourranjbar_Novel_2020}. For example, the authors in \cite{J_Zhang_Joint_2018} considered the problem of joint power and trajectory optimization for an AF relay network consisting of a single UAV as a relay. The problem of using either only one UAV or multiple UAVs as relays is studied in \cite{J_Chen_Multiple_2018}, where the placement of UAVs is optimized by maximizing the end-to-end SNR for both AF and DF relaying protocols. The optimal placement of a UAV-relay for maximum reliability is considered in \cite{J_Chen_Optimum_2018}, where the UAV altitude is also optimized for both static and mobile UAVs. 

{\em Stochastic Geometry for UAV Networks.} Given the irregular locations of transmitters and receivers in modern wireless networks, it is reasonable to consider the random network viewpoint for the system-level analysis of such networks using ideas from stochastic geometry \cite{B_Haenggi_Stochastic_2012, J_Dhillon_Modeling_2012, B_Dhillon_Poisson_2020}. This is particularly relevant for UAV networks, where the UAVs could act either as BSs, UEs, or relays with random placements and movements \cite{B_Morteza_Stochastic_2020, J_Morteza_Handover_2020, J_Amer_Mobility_2020, C_Morteza_3GPP_2019, C_Morteza_Fundamentals_2019, J_Morteza_Impact_2020}. Considering a finite network of UAVs distributed as a binomial point process (BPP), the authors in \cite{J_Chetlur_Downlink_2017} studied the coverage probability of the network for the cases of with and without fading. Motivated by this work, the problem of designing stochastic trajectory processes for mobile UAVs was investigated in \cite{J_Enayati_mobile_2018} and the same coverage trends as in \cite{J_Chetlur_Downlink_2017} were observed. Modeling the locations of UAVs as a BPP, the authors in \cite{J_Wang_Modeling_2018} studied the coexistence problem of UAVs with a network of BSs distributed as a Poisson point process (PPP) on the ground. They have also considered probabilistic LoS/NLoS channel model to further leverage the benefits of using UAVs. Along similar lines, the authors in \cite{J_Alzenad_Coverage_2019} and \cite{J_Cherif_Downlink_2021} analyzed the received rate for a terrestrial and aerial UE, respectively, in a vertical heterogeneous network, comprising of terrestrial BSs and UAVs acting as BSs. Using probabilistic channel model and realistic antenna pattern at the BS site, the problem of finding the optimal spectrum sharing strategy for UAV-to-UAV communications is studied in \cite{J_Azari_UAV_2020}. Considering Poisson cluster processes, the authors in \cite{J_Galkin_Stochastic_2019} investigated the impact of different UAV placement strategies, which could be either independent of or dependent on the UE locations on the ground.

{\em Wirelessly Backhauled UAVs.} Given the growing number of BSs in the forms of terrestrial (macro or small cell) and aerial units, providing strong fiber backhaul for all of these BSs is a challenging task. Therefore, it is inevitable that some of the BSs in a cellular network are wirelessly backhauled to the core network \cite{J_Dhillon_Backhaul_2015, J_Jaber_Wireless_2018, J_Saha_Millimeter_2019}. This is specifically the case for most UAVs, as they are supposed to hover and move freely in the sky, unless being tethered to a building rooftop \cite{J_Kishk_Aerial_2020}. In \cite{C_Kouzayha_Stochastic_2020}, the impact of UAV millimeter-wave (mmWave) backhauling is considered in an aerial-terrestrial cellular network using tools from stochastic geometry. Along similar lines, the authors in \cite{C_Galkin_Backhaul_2018} studied the success probability of establishing a wireless backhaul network using directional antenna patterns for the UAVs. In \cite{J_Sabzehali_Optimizing_2021}, the authors used tools from graph theory to solve a 3D UAV placement problem, where UAVs serve the ground UEs and are also wirelessly backhauled to the terrestrial BSs. Optimal 3D path planning problem for a UAV is investigated in \cite{J_Chowdhury_3D_2020} considering both backhaul constraint and realistic antenna patterns. The main idea was to change the UAV height during the course of its path to improve the backhaul link quality using dynamic programming. Considering link blockages in mmWave frequencies, the authors in \cite{J_Gapeyenko_Flexible_2018} studied the use of UAVs as relays in a flexible backhaul architecture for dynamically rerouting to alternative paths.

Taking full advantage of UAVs for wireless backhaul support requires considering UAV-specific criteria, such as 3D deployment \cite{J_Javad_3D_2021}, realistic antenna patterns (both at the UAV and BS sites) \cite{J_Amer_Toward_2019, C_Amer_Performance_2020}, and a high probability of LoS transmission. In this work, we present the first stochastic geometry-based analysis of 3D UAV-assisted two-hop cellular networks using realistic antenna and channel models. Our key contributions are summarized next.

\vspace{-0.4cm}
\subsection{Contributions} \label{sec:contributions}
This paper provides a comprehensive analysis of downlink transmission in a two-hop UAV-assisted 3D communication system using realistic antenna and channel models. In particular, we model the fiber-backhauled BSs as a 2D homogeneous PPP at a constant height that serve the ground UEs. UAVs are wirelessly backhauled to the BSs and are modeled as a 3D homogeneous PPP hovering between two permissible heights. We assume realistic antenna patterns for the BSs and UAVs based on 3GPP studies \cite{3gpp_36873} and also consider a probabilistic LoS/NLoS channel model for the air-to-ground communication links. Using the max-power association policy for selecting the serving BS and UAV, we consider both AF and DF relaying protocols and adopt a hybrid scheme where a UE is either served directly by a one-hop access link from the serving terrestrial BS or by a two-hop link consisting of an access link from the serving UAV to the UE and a backhaul link from the terrestrial BS to that UAV. Selection between one-hop or two-hop connection is made based on the received SINR \cite{J_Cho_Throughput_2004}. In both cases, we will term the terrestrial BS as the {\em serving BS}, where it serves the UE directly in the one-hop connection and the serving UAV over the backhual link in the two-hop connection. For this setup, we highlight our key contributions next.

\subsubsection{Mathematical Constructs for 3D Relay-Assisted Communication Networks}
We derive the distribution of several random variables that are the building blocks for the analysis of two-hop AF and DF relaying protocols. Furthermore, considering the 3D PPP of UAVs and the probabilistic channel model, we obtain the LoS/NLoS association probabilities and derive the joint distribution of the distance and zenith angle of the serving UAV to the typical UE. We also provide asymptotic results for these distributions.

\subsubsection{Coverage Performance in Backhaul-Aware Communication Networks}
We develop a general framework for analyzing the coverage probability in backhaul-aware two-hop communication networks. For a specific serving UAV channel condition, we derive the conditional Laplace transform of interference for both LoS and NLoS interfering UAVs. Using this Laplace transform along with the distributional results described above, we characterize the coverage probability for both AF and DF relaying protocols.

\subsubsection{Design Insights with Directional Antenna Models}
Inspired by 3GPP documents \cite[Section 5.8]{3gpp_22829}, we propose a novel method to increase the coverage probability of the network by adding a dedicated uptilted directional antenna at the BS site, which is only used for backhaul purposes. This is the first work that considers uptilted antennas at the BS sites for improving aerial coverage. As a baseline, we also consider canonical isotropic antennas, which are vastly used in the literature, and demonstrate the superiority of using directional antennas over them in the simulation results.

\vspace{-0.3cm}
\section{System Model} \label{sec:SysMod}
\vspace{-0.3cm}
\subsection{Spatial Setup} \label{subsec1:SpatialSetup}
\begin{figure}
	\centering
	\includegraphics[width=0.75\columnwidth]{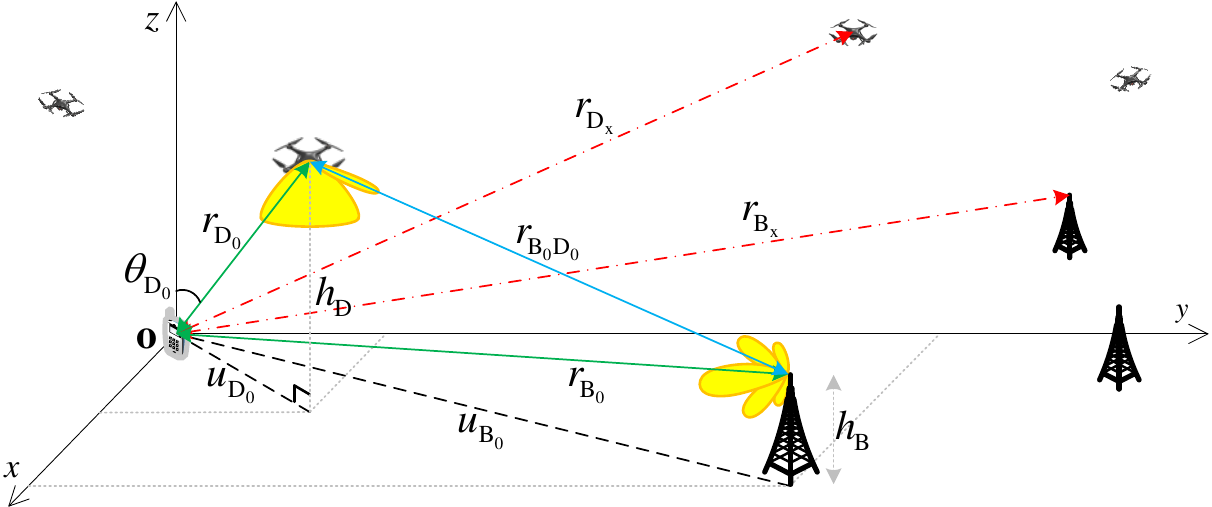}
	\vspace{-0.6cm}
	\caption{An illustration of the system model. Desired and interfering signals to the typical UE are denoted by solid green and dotted red lines, and the access and backhaul links to the typical UE are shown as solid green and blue lines, respectively.}
	\vspace{-0.9cm}
	\label{fig:SystemModel_1}
\end{figure}

We consider a 3D setting where BSs and UAVs coexist to serve UEs on the ground. We assume that BSs have a constant height of $h_{\rm B}$ and the projection of BS locations onto the ground follows a homogeneous PPP $\Phi_{\rm B}$ with density $\lambda_{\rm B}$. Independently from $\Phi_{\rm B}$, UAVs are distributed based on a 3D homogeneous PPP $\Phi_{\rm D}$ with density $\lambda_{\rm D}$ in the space enclosed between heights $h_{\rm D, m}$ and $h_{\rm D, M}$, which represent the minimum and maximum allowable UAV heights, respectively. The ground UEs are distributed as another homogeneous PPP $\Phi_{\rm U}$ independently from $\Phi_{\rm B}$ and $\Phi_{\rm D}$. In this setup, we consider the ground to be aligned with the $xy$-plane of the 3D coordinate system, and without loss of generality, we perform the analysis for the {\em typical UE} placed at the origin $\nbo = (0,0,0)$. As shown in Fig. \ref{fig:SystemModel_1}, we represent the 3D distances from a BS and a UAV located at ${\rm B}_\nbx\in\Phi_{\rm B}$ and ${\rm D}_\nbx\in\Phi_{\rm D}$ to $\nbo$ by $r_{{\rm B}_\nbx} = \sqrt{u_{{\rm B}_\nbx}^2 \!+\! h_{\rm B}^2}$ and $r_{{\rm D}_\nbx} = \sqrt{u_{{\rm D}_\nbx}^2 \!+\! h_{\rm D}^2}$, respectively, where $u_{{\rm B}_\nbx}$ and $u_{{\rm D}_\nbx}$ are the 2D (i.e., horizontal) distances from ${\rm B}_\nbx$ and ${\rm D}_\nbx$ to $\nbo$, respectively. In this paper, we use subscript `$0$' for denoting the serving BS and UAV. Therefore, the locations of the serving BS and UAV are denoted by ${\rm B}_0$ and ${\rm D}_0$, respectively\footnote{With a slight abuse of notation, we represent both the location of the serving BS and the serving BS itself by ${\rm B}_0$. The same goes with ${\rm D}_0$ as well.}, and the 3D and 2D distances from ${\rm B}_0$ (resp. ${\rm D}_0$) to $\nbo$ are denoted by $r_{{\rm B}_0} = \sqrt{u_{{\rm B}_0}^2 \!+\! h_{\rm B}^2}$ and $u_{{\rm B}_0}$ (resp. $r_{{\rm D}_0} = \sqrt{u_{{\rm D}_0}^2 \!+\! h_{{\rm D}_0}^2}$ and $u_{{\rm D}_0}$), respectively, where $h_{{\rm D}_0}$ is the serving UAV height. We represent the 3D distance between ${\rm B}_0$ and ${\rm D}_0$ by $r_{{\rm B}_0{\rm D}_0}$. More details on the serving BS and UAV are given in Section \ref{subsec4:AssociationPolicy}.

\begin{remark}
	Since the UAVs should be able to hover at any location, we require them to be rotary-wing drones in this paper.
\end{remark}

\vspace{-0.8cm}
\subsection{Antenna Pattern} \label{subsec2:AntennaPattern}
We adopt realistic antenna radiation patterns \cite{3gpp_36873} for BSs, UAVs, and UEs, as explained next.




\subsubsection{BS}
We consider two different antenna models at the BSs: (i) omnidirectional antenna\footnote{In antenna theory terminology, ``omnidirectional" refers to constant radiation pattern only in the \textit{horizontal} direction \cite{B_Stutzman_Antenna_2012}.}, and (ii) a combination of omnidirectional and directional antennas. Although our main focus in this paper is on realistic antenna patterns, we will also study the canonical isotropic antenna pattern, which has the same radiation pattern in \textit{all} directions, as a baseline for comparison.



\paragraph{Downtilted Omnidirectional}\label{par:OmniBS}
In this model proposed by 3GPP, we consider a uniform linear array (ULA) that is vertically installed at each BS. The ULA has $N_{\rm B}$ elements, equally separated by $\lambda/2$, where $\lambda = {\rm c}/f$ is the wavelength of the operating frequency $f$ and ${\rm c}$ is the speed of light. The \textit{normalized array factor} for this ULA can be written as \cite[Sec. 8.3.2]{B_Stutzman_Antenna_2012}
\begin{align}\label{G_ArrayFactor}
f_{\rm A}(\theta, \theta_{\rm B}) = \frac{\sin\left(N_{\rm B} \frac{\pi}{2}[\cos(\theta) - \cos(\theta_{\rm B})]\right)}{N_{\rm B}\sin\left(\frac{\pi}{2}[\cos(\theta) - \cos(\theta_{\rm B})]\right)},
\end{align}
where $\theta$ and $\theta_{\rm B}$ are, respectively, the zenith angle and the direction of the BS antenna mainlobe, both measured from the $z$-axis of the 3D coordinate system. Note that since the primary objective of terrestrial BSs is to serve the ground UEs, their antenna mainlobes are tilted downward to the ground, which means that $\pi/2 < \theta_{\rm B} < \pi$. Each element of the ULA is an omnidirectional antenna that has a normalized vertical radiation power pattern (in dB) of $G_{\rm E, V}(\theta) = -\min\left\{ 12\left(\frac{\theta - \pi/2}{\theta_{\rm 3dB}}\right)^2,~ {\rm SLA_V} \right\}$, 
where $\theta_{\rm 3dB} = 65^\circ$ is the vertical $3$dB beamwidth and ${\rm SLA_V} = 30$ dB is the sidelobe attenuation limit \cite{3gpp_36873}. Note that this antenna is omnidirectional along the horizontal direction with a normalized gain of $G_{\rm E, H}(\phi) = 0$ dB, where $\phi$ is the azimuthal angle measured from the $x$-axis in the $xy$-plane. Furthermore, we assume the maximum gain of each antenna element is $G_{\rm E}^{\max} = 8$ dBi \cite{3gpp_36873}. Therefore, we write the \textit{3D element pattern} as $G_{\rm E, 3D}(\theta, \phi) = G_{\rm E}^{\max} + G_{\rm E, V}(\theta) + G_{\rm E, H}(\phi) = G_{\rm E}^{\max} + G_{\rm E, V}(\theta) := G_{\rm E}(\theta)$.
The complete gain of the BS antenna array along direction $\theta$ for all $\phi$ can now be written in dBi as \cite[Sec. 8.4]{B_Stutzman_Antenna_2012}
\begin{align}\label{G_OmniD}
G_{\rm B}^{\rm OmniD}(\theta, \theta_{\rm B}) = G_{\rm E}(\theta) + 20\log(\left|f_{\rm A}(\theta, \theta_{\rm B})\right|),
\end{align}
where the superscript ${\rm OmniD}$ stands for ``downtilted omnidirectional".

\paragraph{Downtilted Omnidirectional and Uptilted Directional} \label{par:OmniDirBS}
Looking closely at the previous model, we observe that UAVs are mostly served by the BS antenna sidelobes, which is a major drawback of this model since sidelobe peak gain is usually much lower than the mainlobe peak gain. For instance, the sidelobe level for a ULA with a large number of antenna elements is about $-13.3$ dB \cite[Sec. 8.3.1]{B_Stutzman_Antenna_2012}. Furthermore, UAVs may be in the null direction of the BS antenna array, which could cause an outage in the UAV-BS link. As discussed in 3GPP TR 22.829 \cite[Section 5.8]{3gpp_22829}, one way to combat these effects and improve the backhaul link is to deploy a separate \textit{uptilted} directional antenna along with the aforementioned downtilted omnidirectional antenna array at the BS. Note that this newly added directional antenna is merely used for backhaul purposes, i.e., communicating with the UAVs. It is worth mentioning that while this work was under review, other researchers have also used the idea of adding equipment at the BS sites to better accommodate UAVs, see e.g., \cite{J_Maeng_Base_2021, J_Chowdhury_Ensuring_2021}. In both these works, the authors try to optimize the tilt angle of the BS uptilted antenna to avoid aerial coverage holes and guarantee reliable communications.


Similar to the downtilted omnidirectional antenna model, we use the 3GPP-based antenna pattern here as well, which can be either mechanically or electrically steered toward the UAV locations. More specifically, when directed toward $(\theta_0, \phi_0)$, this antenna has normalized vertical, horizontal, and 3D radiation power patterns (in dB) of $G_{\rm V}(\theta, \theta_0) = -\min\{ 12(\frac{\theta - \theta_0}{\theta_{\rm 3dB}})^2,~ {\rm SLA_V} \}$, $G_{\rm H}(\phi, \phi_0) = -\min\{ 12(\frac{\phi - \phi_0}{\phi_{\rm 3dB}})^2,~ A_{\rm m} \}$, and $G_{\rm 3D}(\theta, \phi, \theta_0, \phi_0) = -\min\left\{ -G_{\rm V}(\theta, \theta_0) - G_{\rm H}(\phi, \phi_0),~ A_{\rm m} \right\}$,
respectively, where $\theta_{\rm 3dB} = 10^\circ$ and $\phi_{\rm 3dB} = 10^\circ$ are the vertical and horizontal $3$dB beamwidths and $A_{\rm m} = 30$ dB is the front-back ratio \cite{3gpp_36873}. Note that we used narrower beamwidths than those in the downtilted omnidirectional antenna to reduce interference. Furthermore, since this antenna is tilted upward, we require that $0 < \theta_0 < \pi/2$. Hence, the gain of this antenna along direction $(\theta, \phi)$ in dBi becomes $G_{\rm B}^{\rm DirU}(\theta, \phi, \theta_0, \phi_0) = G^{\max} + G_{\rm 3D}(\theta, \phi, \theta_0, \phi_0)$,
where $G^{\max} = 8$ dBi is the maximum gain of this antenna and the superscript ${\rm DirU}$ stands for ``uptilted directional". 



\subsubsection{UAV}


We assume UAVs are equipped with two sets of antennas, one for the backhaul connection and the other for the access link (see Fig. \ref{fig:SystemModel_1}). The backhaul antenna is directional and has the same pattern as the uptilted directional antenna at the BS described earlier in Section \ref{par:OmniDirBS}, with the only difference that its main beam is not necessarily tilted upward. In fact, since the height of the UAVs are usually higher than that of the BSs, the UAV backhaul antenna is usually downtilted. Hence, we write the gain of this antenna as $G_{\rm D}^{\rm BH}(\theta, \phi, \theta_0, \phi_0) = G_{\rm B}^{\rm DirU}(\theta, \phi, \theta_0, \phi_0)$,
where $\pi / 2 < \theta_0 < \pi$ and the superscript ${\rm BH}$ stands for ``backhaul". On the other hand, the access antenna is assumed to be downtilted omnidirectional with the following features: (i) static (non-steerable), so the mainlobe direction cannot change, (ii) tilted completely toward the ground with $\theta_0 = \pi$, and (iii) has a wider beam than the backhaul antenna to serve the UEs and we set $\theta_{\rm 3dB}= 120^\circ$. Thus, the UAV access antenna gain can be written in dBi as
\begin{align}\label{G_UAV}
G_{\rm D}^{\rm AC}(\theta) &= G^{\max} - \min\left\{ 12\left(\frac{\theta - \pi}{\theta_{\rm 3dB}}\right)^2,~ {\rm SLA} \right\},
\end{align}
where $G^{\max} = 8$ dBi, ${\rm SLA} = 30$ dB \cite{3gpp_36873}, and the superscript ${\rm AC}$ stands for ``access". 


\subsubsection{UE}
Each UE is equipped with an isotropic antenna with gain $G_{\rm U} = 0$ dBi in all directions.
\vspace{-0.8cm}
\begin{remark}
As mentioned earlier, the antenna mainlobe directions for both the BS omnidirectional and the UAV access antennas are static, while the BS directional antenna (which is used only for backhaul) and the UAV backhaul antenna have both steerable mainlobe directions. As for the UAV and the terrestrial BS to which it is backhauled, we assume that their backhaul antennas are steered exactly toward each other, while the antenna directions of interfering BSs and UAVs are chosen uniformly at random.
\end{remark}

\vspace{-0.7cm}
\subsection{Channel Model} \label{subsec3:ChannelModel}

\subsubsection{LoS and NLoS Channel Conditions} \label{subsubsec3.2:LoS}
One of the major advantages of employing UAVs in wireless communications is their superior channel conditions as compared to their terrestrial counterparts. In fact, since UAVs usually fly at high altitudes, they are expected to have a high probability of LoS, which results in low attenuation in the received signal \cite{3gpp_36777}. To capture this unique feature of aerial networks, we consider a mixture of LoS and NLoS links for the channel model. Specifically, a UAV establishes an LoS link with the typical UE with probability \cite{J_AlHourani_Optimal_2014}
\begin{align}\label{eq:LoS}
p_{\rm L}(\theta) = \frac{1}{1 + c_1 {\rm e}^{-c_2(90 - \theta - c_1)}},
\end{align}
and an NLoS link with probability $p_{\rm N}(\theta) = 1 - p_{\rm L}(\theta)$, where $\theta$ is the zenith angle measured from the $z$-axis in degrees ($90 - \theta$ is the elevation angle), and $c_1$ and $c_2$ are two positive environment-dependent parameters. From \eqref{eq:LoS}, we observe that as the UAV height increases, $\theta$ decreases, and thus, $p_{\rm L}(\theta)$ will increase. Note that we use this model only for the UAV-UE channels. Since the BS heights are usually comparable to those of buildings in urban or rural environments, we assume that BS-UAV and BS-UE links are always in LoS and NLoS conditions, respectively.

\subsubsection{Received Powers} \label{subsubsec3.1:ReceivedPowers}
We assume that BSs and UAVs transmit with constant powers $P_{\rm B}$ and $P_{\rm D}$, respectively. For consistency, we represent the antenna gains of the BSs and UAVs toward the typical UE (access) by $G$ and toward each other and among themselves (backhaul) by $g$. Since the BS-UE link is in an NLoS condition, the received power at the typical UE from the serving BS can be written as $P_{{\rm B}_0}^{\rm Rx} = P_{\rm B} G_{{\rm B}_0} G_{\rm U} f_{{\rm B}_0} r_{{\rm B}_0}^{-\alpha_{\rm N}} \eta_{\rm N}^{-1}$, where $G_{{\rm B}_0}$ is the serving BS antenna gain along the direction of the typical UE, $G_{\rm U}$ is the typical UE antenna gain, $f_{{\rm B}_0}$ is the small-scale fading power between the serving BS and the typical UE, and $\alpha_{\rm N}$ and $\eta_{\rm N}$ are the path-loss exponent and the mean excessive path-loss for NLoS transmission, respectively \cite{J_AlHourani_Optimal_2014}. On the other hand, the BS-UAV link is in an LoS condition, and thus, we write the received power at the serving UAV from the serving BS as $P_{{\rm B}_0{\rm D}_0}^{\rm Rx} = P_{\rm B} g_{{\rm B}_0} g_{{\rm D}_0} f_{{\rm B}_0{\rm D}_0} r_{{\rm B}_0{\rm D}_0}^{-\alpha_{\rm L}} \eta_{\rm L}^{-1}$, where $g_{{\rm B}_0}$ is the serving BS antenna gain along the direction of the serving UAV, $g_{{\rm D}_0}$ is the serving UAV antenna gain along the direction of the serving BS, $f_{{\rm B}_0{\rm D}_0}$ is the small-scale fading power between the serving BS and the serving UAV, and $\alpha_{\rm L}$ and $\eta_{\rm L}$ are the path-loss exponent and the mean excessive path-loss for LoS transmission, respectively. Since the UAV-UE link may experience both channel conditions, we write the received power at the typical UE from the serving UAV for the LoS and NLoS conditions, respectively, as $P_{\rm D_0, L}^{\rm Rx} = P_{\rm D} G_{{\rm D}_0} G_{\rm U} f_{{\rm D}_0} r_{\rm D_0, L}^{-\alpha_{\rm L}} \eta_{\rm L}^{-1}$ and $P_{\rm D_0, N}^{\rm Rx} = P_{\rm D} G_{{\rm D}_0} G_{\rm U} f_{{\rm D}_0} r_{\rm D_0, N}^{-\alpha_{\rm N}} \eta_{\rm N}^{-1}$, where $G_{{\rm D}_0}$ is the serving UAV antenna gain along the direction of the typical UE, $f_{{\rm D}_0}$ is the small-scale fading power between the serving UAV and the typical UE, and $r_{\rm D_0, L}$ and $r_{\rm D_0, N}$ are the serving UAV distances to the typical UE in LoS and NLoS conditions, respectively. Note that we have $\alpha_{\rm L} < \alpha_{\rm N}$ and $\eta_{\rm L} < \eta_{\rm N}$.

Let us now define interference at the typical UE and the serving UAV, which are of interest for the downlink analysis. We represent the set of interfering BSs and UAVs by $\Phi_{\rm B}' \coloneqq \Phi_{\rm B}\backslash {\rm B}_0$ and $\Phi_{\rm D}' \coloneqq \Phi_{\rm D}\backslash {\rm D}_0$, respectively, and write the received power at the typical UE and the serving UAV from the interfering BSs ($I_{\rm BU}$ and $I_{\rm BD}$) and the received power at the serving UAV from the interfering UAVs ($I_{\rm DD}$) as $I_{\rm BU} = \sum_{{\rm B}_\nbx \in \Phi_{\rm B}'} P_{\rm B} G_{{\rm B}_\nbx} G_{\rm U} f_{{\rm B}_\nbx} r_{{\rm B}_\nbx}^{-\alpha_{\rm N}} \eta_{\rm N}^{-1}$, $I_{\rm BD} = \sum_{{\rm B}_\nbx \in \Phi_{\rm B}'} P_{\rm B} g_{{\rm B}_\nbx} g_{{\rm D}_0} f_{{\rm B}_\nbx{\rm D}_0} r_{{\rm B}_\nbx{\rm D}_0}^{-\alpha_{\rm L}} \eta_{\rm L}^{-1}$, and $I_{\rm DD} = \sum_{{\rm D}_\nbx \in \Phi_{\rm D}'} P_{\rm D} g_{{\rm D}_\nbx} g_{{\rm D}_0} f_{{\rm D}_\nbx{\rm D}_0} r_{{\rm D}_\nbx{\rm D}_0}^{-\alpha_{\rm L}} \eta_{\rm L}^{-1}$, respectively, where the parameters $G_{{\rm B}_\nbx}$, $f_{{\rm B}_\nbx}$, $g_{{\rm B}_\nbx}$, $f_{{\rm B}_\nbx{\rm D}_0}$, $g_{{\rm D}_\nbx}$, and $f_{{\rm D}_\nbx{\rm D}_0}$ are defined similarly as in the serving BS/UAV parameters described earlier. As for the received power at the typical UE from the interfering UAVs, we first partition the set of all UAVs in two disjoint sets of LoS ($\Phi_{\rm D, L}$) and NLoS ($\Phi_{\rm D, N}$) UAVs. Using thinning theorem for the PPP $\Phi_{\rm D}$, we observe that $\Phi_{\rm D, L}$ and $\Phi_{\rm D, N}$ are two independent inhomogeneous PPPs with densities $\lambda_{\rm D} p_{\rm L}(\theta)$ and $\lambda_{\rm D} p_{\rm N}(\theta)$, respectively \cite{B_Haenggi_Stochastic_2012}. The point process of interfering UAVs in LoS and NLoS conditions can now be defined as $\Phi_{\rm D, L}' \coloneqq \Phi_{\rm D, L}\backslash {\rm D}_0$ and $\Phi_{\rm D, N}' \coloneqq \Phi_{\rm D, N}\backslash {\rm D}_0$. Finally, we define the received power at the typical UE from each set as $I_{\rm DU, L} = \sum_{{\rm D}_\nbx \in \Phi_{\rm D, L}'} P_{\rm D} G_{{\rm D}_\nbx} G_{\rm U} f_{{\rm D}_\nbx} r_{{\rm D}_\nbx}^{-\alpha_{\rm L}} \eta_{\rm L}^{-1}$ and $I_{\rm DU, N} = \sum_{{\rm D}_\nbx \in \Phi_{\rm D, N}'} P_{\rm D} G_{{\rm D}_\nbx} G_{\rm U} f_{{\rm D}_\nbx} r_{{\rm D}_\nbx}^{-\alpha_{\rm N}}\eta_{\rm N}^{-1}$, where the parameters $G_{{\rm D}_\nbx}$ and $f_{{\rm D}_\nbx}$ are defined similarly as before. We also represent the total interference from the UAVs at the typical UE by $I_{\rm DU} = I_{\rm DU, L} + I_{\rm DU, N}$, and the total interference from both the BSs and UAVs at the typical UE by $I_{\rm U} = I_{\rm BU} + I_{\rm DU}$.
The received SINR at the typical UE from the serving BS, the received SINR at the typical UE from the serving UAV, and the received SINR at the serving UAV from the serving BS are defined, respectively, as
\begin{align}\label{SINR}
{{\rm SINR}_{{\rm BU}, q}} \!=\! \frac{P_{{\rm B}_0}^{\rm Rx}}{I_{\rm U} \!+\! P_{{\rm D_0}, q}^{\rm Rx} \!+\! N_0}, \quad {{\rm SINR}_{{\rm DU}, q}} \!=\! \frac{P_{{\rm D_0}, q}^{\rm Rx}}{I_{\rm U} \!+\! P_{{\rm B}_0}^{\rm Rx} \!+ \! N_0}, \quad {\rm SINR_{BD}} \!=\! \frac{P_{{\rm B}_0{\rm D}_0}^{\rm Rx}}{I_{\rm BD} \!+\! I_{\rm DD} \!+\! N_0},
\end{align}
where $q = \{\rm L, N\}$ denotes the LoS or NLoS channel conditions and $N_0$ is the noise power.

\begin{assumption}\label{Assumption1}
For \emph{non-isotropic} antennas, the total interference at the serving UAV from other BSs ($I_{\rm BD}$) and UAVs ($I_{\rm DD}$) is negligible and assumed to be $0$ in this paper. Therefore, we have ${\rm SINR_{BD}} \approx {\rm SNR_{BD}}= \frac{P_{{\rm B}_0{\rm D}_0}^{\rm Rx}}{N_0}$, where ${\rm SNR_{BD}}$ is the SNR at the serving UAV from the serving BS.
\end{assumption}


\subsubsection{Fading} \label{subsubsec3.3:Fading}
We consider Nakagami-$m$ fading model for both the LoS and NLoS channels since it captures a wide variety of fading environments\footnote{The most natural choice for modeling small-scale fading in UAV-assisted communications is Rician fading, which makes a clear distinction between the direct and scattered paths. This is mainly due to the high probability of LoS in aerial networks \cite{J_AlHourani_Optimal_2014}. However, the Rician pdf does not lend itself to further analysis since it entails modified Bessel function. Because of this, we use the Nakagami-$m$ fading model instead of the Rician model, which is quite common in the literature for system-level analysis thanks to its  mathematical tractability. Furthermore, using the moment matching technique, it is well-known that the Rician distribution with factor $K$ can be well approximated with the Nakagami-$m$ distribution using the relation $m = \frac{(K+1)^2}{2K+1}$.}. Hence, the channel fading powers $f_{{\rm B}_0}$, $f_{{\rm D}_0}$, $f_{{\rm B}_0{\rm D}_0}$, $f_{{\rm B}_\nbx}$, $f_{{\rm D}_\nbx}$, $f_{{\rm B}_\nbx{\rm D}_0}$, and $f_{{\rm D}_\nbx{\rm D}_0}$ are all gamma distributed with probability density function (pdf) and cumulative distribution function (cdf) of $f_X(x) = \frac{m ^ m}{\Gamma(m)} x ^ {m - 1} {\rm e}^{-m x}$ and $F_x(x) = \frac{1}{\Gamma(m)}\gamma(m, mx)$, respectively, where $\gamma(s, x) = \int_0^x t^{s-1} {\rm e}^{-t}\,{\rm d}t$ is the lower incomplete gamma function and $\Gamma(s) = \gamma(s, \infty)$ is the gamma function. For mathematical tractability, we assume that $m$ is integer and the serving and interfering links have the same $m$ values.

\vspace{-0.5cm}
\subsection{Service Model and Association Policy} \label{subsec4:AssociationPolicy}
We assume BSs have strong and reliable fiber backhaul connections to the core network, while UAVs are wirelessly backhauled to the BSs. To connect UEs to the core network, we consider the following two service models: (i) access only, where the UEs connect directly to the BSs via access links, and (ii) joint access and backhaul, where the UEs use UAVs as relays for connecting to the BSs. In the second service model, UAV-UE and BS-UAV links are regarded as the access and backhaul links, respectively. In this paper, we use a \textit{hybrid} scheme where the UEs can be served either directly by the BS-UE links or indirectly by a two-hop connection consisting of the BS-UAV and UAV-UE links. The selection between the one-hop and two-hop connections is made based on the SINR \cite{J_Cho_Throughput_2004}. Using the maximum received power association policy, we write the association rules as
\vspace{-0.2cm}
\begin{align}
{\rm B}_0 &=\underset{{\rm B}_\nbx\in\Phi_{\rm B}}{\argmax} ~ P_{{\rm B}_\nbx}^{\rm Rx} = \underset{{\rm B}_\nbx\in\Phi_{\rm B}}{\argmax} ~ r_{{\rm B}_\nbx}^{-\alpha_{\rm N}} = \underset{{\rm B}_\nbx\in\Phi_{\rm B}}{\argmin} ~ r_{{\rm B}_\nbx},\label{Assoc_BS}\\
{\rm D}_0 &= \underset{q\in\{{\rm L, N}\},~{\rm D}_\nbx\in\Phi_{{\rm D}, q}}{\argmax} P_{{\rm D}_\nbx, q}^{\rm Rx} = \underset{q\in\{{\rm L, N}\},~{\rm D}_\nbx\in\Phi_{{\rm D}, q}}{\argmax} r_{{\rm D}_\nbx}^{-\alpha_q} \eta_{q}^{-1},\label{Assoc_UAV}
\end{align}
where the impact of antenna gains is absorbed into $\eta_{q}$ for simplicity. Note that since all the BSs experience the NLoS channel condition when connecting to the typical UE, the maximum average received power and the nearest neighbor association policies are equivalent for the BSs. However, this is not the case for the UAVs, as a farther UAV to the typical UE may have better channel conditions than a nearer one, and thus, be regarded as the serving UAV.


\vspace{-0.4cm}
\subsection{Relaying Protocols and Metrics} \label{subsec5:RelayingMetric}
We adopt both AF and DF relaying protocols in this paper. Considering AF downlink transmission, the received signal from the source (BS) at the relay (UAV) is multiplied by a gain $G$ before being forwarded to the destination (UE). Note that the choice of the relay gain $G$ defines the overall performance of the AF relaying protocol \cite{J_Hasna_End_2003, J_Laneman_Cooperative_2004}. Assuming $G = 1 / (P_{{\rm B}_0{\rm D}_0}^{\rm Rx} + I_{\rm BD} + I_{\rm DD} + N_0)$, the AF end-to-end SINR for the UAV-UE channel condition $q$ can be written as
\begin{align}\label{eq:AF_SINR}
{{\rm SINR}_{{\rm e2e}, q}^{\rm AF}} = \frac{{\rm SINR_{BD}} {{\rm SINR}_{{\rm DU}, q}}}{{\rm SINR_{BD}} + {{\rm SINR}_{{\rm DU}, q}} + 1}.
\end{align}
In the DF protocol, the received signal is first decoded and then forwarded to the destination. Since both the relay and destination nodes must decode the source signal without error for a successful transmission \cite{J_Laneman_Cooperative_2004}, we write the DF end-to-end SINR for channel condition $q$ as
\vspace{-0.1cm}
\begin{align}\label{eq:DF_SINR}
{{\rm SINR}_{{\rm e2e}, q}^{\rm DF}} = \min\left\{ {\rm SINR_{BD}}, {{\rm SINR}_{{\rm DU}, q}} \right\}.
\end{align}
We can easily show that the DF protocol always outperforms the AF protocol. In fact, we have ${{\rm SINR}_{{\rm e2e}, q}^{\rm DF}} > \left( {\rm SINR_{BD}}^{-1} + {{\rm SINR}_{{\rm DU}, q}}^{-1} \right)^{-1} > {{\rm SINR}_{{\rm e2e}, q}^{\rm AF}}$. We represent the received SINR at the typical UE by ${\rm SINR^{AF}}$ and ${\rm SINR^{DF}}$ for the AF and DF relaying protocols, respectively, and define them for specific channel condition $q$ using the aforementioned hybrid scheme as
\begin{align}\label{eq:SINR}
{\rm SINR}_q^{\rm AF} = \max\left\{ {\rm SINR}_{{\rm BU}, q}, {\rm SINR}_{{\rm e2e}, q}^{\rm AF} \right\}, \qquad{\rm SINR}_q^{\rm DF} = \max\left\{ {\rm SINR}_{{\rm BU}, q}, {\rm SINR}_{{\rm e2e}, q}^{\rm DF} \right\}.
\end{align}
To evaluate the network performance, we introduce coverage probability as our main metric, which is defined as the probability that the received SINR at the typical UE exceeds a predetermined constant threshold $\tau$, i.e., $P_{\rm Cov}^{\rm AF} = {\mathbb P}[{\rm SINR^{AF}} \geq \tau]$ and $P_{\rm Cov}^{\rm DF} = {\mathbb P}[{\rm SINR^{DF}} \geq \tau]$ for the AF and DF relaying protocols, respectively.


\vspace{-0.4cm}
\section{Mathematical Constructs} \label{sec:MathematicalConstructs}
In this section, we provide some important intermediate results that help us analyze the coverage probability in 3D UAV-assisted communication networks.
\vspace{-0.5cm}
\subsection{Useful Lemmas for the Two-Hop Setting} \label{sec:TwoHopSetting}
Conditioned on knowing $\rm B_0$ and $\rm D_0$, one can represent the SINR values given in \eqref{SINR} as
\begin{align}\label{SINR2}
{\rm SINR}_{{\rm BU}, q} = \frac{aX}{bY + I}, \qquad {\rm SINR}_{{\rm DU}, q} = \frac{bY}{aX + I}, \qquad {\rm SINR_{BD}} = \frac{cZ}{N_0},
\end{align}
where $a = P_{\rm B} G_{{\rm B}_0} r_{{\rm B}_0}^{-\alpha_{\rm N}} \eta_{\rm N}^{-1}$, $b = P_{\rm D} G_{{\rm D}_0} r_{{\rm D_0}, q}^{-\alpha_q} \eta_q^{-1}$, $c = P_{\rm B} g_{{\rm B}_0} g_{{\rm D}_0} r_{{\rm B}_0{\rm D}_0}^{-\alpha_{\rm L}} \eta_{\rm L}^{-1}$, $I = I_{\rm U} + N_0$, $X = f_{{\rm B}_0}$, $Y = f_{{\rm D}_0}$, $Z = f_{{\rm B}_0{\rm D}_0}$, and we used Assumption \ref{Assumption1} in writing ${\rm SINR_{BD}}$. Since we assumed the Nakagami-$m$ fading model, $X$, $Y$, and $Z$ are distributed as gamma random variables. In the next three lemmas, we will characterize the statistics of some functions of these gamma random variables that are useful for analyzing the performance of UAV-assisted two-hop relay networks.
\begin{lemma}\label{lem:01}
Let $X$ and $Y$ be two independent gamma random variables with integer-valued shape and rate parameters both equal to $m$, and let $a$, $b$, and $I$ be given non-negative constants. Then the cdf of $T_1 = \frac{aX}{bY + I}$ can be written as
\begin{align} \label{CDF1}
F_{T_1}(\tau) = 1 - \sum_{i=0}^{m-1}\sum_{k=0}^{i} \frac{(k+m-1)!}{k!(m-1)!(i-k)!} \frac{a^m (b\tau)^k}{(a + b\tau)^{m + k}} \left(\frac{m\tau}{a}I\right)^{i-k} {\rm e}^{-\frac{m\tau}{a}I}.
\end{align}
\end{lemma}
\vspace{-0.2cm}
\begin{IEEEproof}
See Appendix \ref{app:lem:01}.
\end{IEEEproof}
\vspace{-0.4cm}
\begin{lemma}\label{lem:02}
Let $X$ and $Y$ be two independent gamma random variables with integer-valued shape and rate parameters both equal to $m$, and let $a$, $b$, and $I$ be given non-negative constants. Then the cdf of $T_2 = \frac{\max\{aX, bY\}}{\min\{aX, bY\} + I}$ can be written as
\begin{align} \label{CDF2}
F_{T_2}(\tau) &= \sum_{i=0}^{m-1} \frac{\gamma\left(m+i, \left(\frac{1}{a} + \frac{1}{b}\right)\frac{m\tau}{|1-\tau|{\bf 1}(\tau < 1)}I\right)}{i!(m-1)!} \frac{a^m b^i + a^i b^m}{(a + b)^{m + i}} \nonumber\\
&\quad-\sum_{i=0}^{m-1}\sum_{k=0}^{i} \frac{\gamma\left(m+k, \left(\frac{\tau}{a} + \frac{1}{b}\right)\frac{m\tau}{|1-\tau|{\bf 1}(\tau < 1)}I\right)}{k!(m-1)!(i-k)!} \frac{a^m (b\tau)^k}{(a + b\tau)^{m + k}} \left(\frac{m\tau}{a}I\right)^{i-k} {\rm e}^{-\frac{m\tau}{a}I} \nonumber\\
&\quad-\sum_{i=0}^{m-1}\sum_{k=0}^{i} \frac{\gamma\left(m+k, \left(\frac{1}{a} + \frac{\tau}{b}\right)\frac{m\tau}{|1-\tau|{\bf 1}(\tau < 1)}I\right)}{k!(m-1)!(i-k)!} \frac{(a\tau)^k b^m}{(a\tau + b)^{k + m}} \left(\frac{m\tau}{b}I\right)^{i-k} {\rm e}^{-\frac{m\tau}{b}I},
\end{align}
where ${\bf 1}(.)$ is the indicator function.
\end{lemma}
\vspace{-0.4cm}
\begin{IEEEproof}
See Appendix \ref{app:lem:02}.
\end{IEEEproof}
The following lemma characterizes a joint cdf that will be used in analyzing the coverage probability in the AF relaying protocol.
\vspace{-0.3cm}
\begin{lemma}\label{lem:03}
Let $X$ and $Y$ be two independent gamma random variables with integer-valued shape and rate parameters both equal to $m$, and let $a$, $b$, $I$, and $g$ be given non-negative constants. Then the joint cdf of $T_1 = \frac{aX}{bY + I}$ and $T_3 = \frac{bY}{aX + I + g(aX + bY + I)}$ when $\tau < \frac{1}{g}$ can be written as
\begin{align} \label{CDF3}
F_{T_1, T_3}(\tau, \tau) &= \sum_{i=0}^{m-1} \frac{\gamma\left(m+i, \left(\frac{1}{a(1+g)} + \frac{1}{b}\right)\frac{m\tau (1+g)}{|1-\tau (1+g)|{\bf 1}(\tau < \frac{1}{1+g})}I\right)}{i!(m-1)!} \frac{(a(1+g))^m b^i + (a(1+g))^i b^m}{(a(1+g) + b)^{m + i}} \nonumber\\
&\quad-\sum_{i=0}^{m-1}\sum_{k=0}^{i} \frac{\gamma\left(m+k, \left(\frac{\tau}{a} + \frac{1}{b}\right)\frac{m\tau (1+g)}{|1-\tau (1+g)|{\bf 1}(\tau < \frac{1}{1+g})}I\right)}{k!(m-1)!(i-k)!} \frac{a^m (b\tau)^k}{(a + b\tau)^{m + k}} \left(\frac{m\tau}{a}I\right)^{i-k} {\rm e}^{-\frac{m\tau}{a}I} \nonumber\\
&\quad-\sum_{i=0}^{m-1}\sum_{k=0}^{i} \frac{\gamma\left(m+k, \left(\frac{1}{a(1+g)} + \frac{\tau}{b(1-\tau g)}\right) \frac{m\tau (1+g)}{|1-\tau (1+g)|{\bf 1}(\tau < \frac{1}{1+g})}I\right)}{k!(m-1)!(i-k)!} \nonumber\\
&\qquad\times\frac{(a\tau (1+g))^k (b(1-\tau g))^m}{(a\tau (1+g) + b(1-\tau g))^{k + m}} \left(\frac{m\tau (1+g)}{b(1-\tau g)}I\right)^{i-k} {\rm e}^{-\frac{m\tau (1+g)}{b(1-\tau g)}I},
\end{align}
and when $\tau \geq \frac{1}{g}$, we have $F_{T_1, T_3}(\tau, \tau) = F_{T_1}(\tau)$, where $F_{T_1}(\tau)$ is given in Lemma \ref{lem:01}.
\end{lemma}
\vspace{-0.3cm}
\begin{IEEEproof}
See Appendix \ref{app:lem:03}.
\end{IEEEproof}
\vspace{-0.3cm}
\begin{remark}\label{remark5}
From the previous lemmas, we observe the following special cases:
\begin{itemize}
\item $\tau = 0$. All the cdfs tend to $0$. For $T_1$, all the terms in the double summation are $0$, except for $i=k=0$, which is $1$, making $F_{T_1}(0) = 0$. For $T_2$, since $\gamma(s, 0) = 0$ for all $s$, we have $F_{T_2}(0) = 0$. The same reasoning applies to the joint cdf of $T_1$ and $T_3$, giving $F_{T_1, T_3}(0, 0) = 0$.
\item $\tau\to \infty$. All the cdfs tend to $1$. For $T_1$, the double summation will be $0$, making $F_{T_1}(\infty) = 1$. For $T_2$, note that the double summations are both $0$, while the single summation is equal to $1$. This can be shown as follows:
\begin{align}\label{eq:rem5}
\sum_{i=0}^{m-1} \! {m\!+\!i\!-\!1 \choose i} &\frac{a^m b^i + a^i b^m}{(a + b)^{m+i}} = \frac{1}{(a + b)^{2m-1}}\sum_{i=0}^{m-1} \!{m\!+\!i\!-\!1 \choose i} (a^m b^i + a^i b^m)(a+b)^{m-1-i}\nonumber\\
&=\frac{\sum_{i=0}^{m-1}\sum_{k=0}^{i} {2(m-1)-i \choose m-1}{i \choose k} (a^{m+k} b^{m-1-k} + a^{m-1-i+k} b^{m+i-k})}{\sum_{l=0}^{2m-1} {2m-1 \choose l} a^l b^{2m-1-l}},
\end{align}
where in the last equality we expanded $(a+b)^{m-1-i}$ and used the change of variables $m-1-i \mapsto i$ in the numerator. To prove the last equation is unity, we need to show that the coefficients of $a^l b^{2m-1-l}$ for $0 \leq l \leq 2m - 1$ are equal in the numerator and denominator of \eqref{eq:rem5}, which can be verified using the following binomial identity \cite[Eq. (1.78)]{B_Gould_Combinatorial_2010}:
\begin{align*}
\sum_{k=0}^{n} {\alpha+k \choose k} {r+n-k \choose n-k} = {\alpha+r+n+1 \choose n}, \qquad \forall \alpha, r.
\end{align*}
Hence, $F_{T_2}(\infty) = 1$. As for the joint cdf of $T_1$ and $T_3$, we have $F_{T_1, T_3}(\infty, \infty) = F_{T_1}(\infty) = 1$.
\item $g = 0$. The joint cdf of $T_1$ and $T_3$ is equivalent to the cdf of $T_2$.
\item $g \to \infty$. The joint cdf of $T_1$ and $T_3$ is equivalent to the cdf of $T_1$, since $T_3 \to 0$.
\end{itemize}
\end{remark}
In case of Rayleigh fading, the fading powers will have exponential distribution, which is a special case of the gamma distribution. Corollary \ref{cor:1} gives the results of the previous lemmas for Rayleigh fading, which has a straightforward proof by setting $m = 1$ in \eqref{CDF1}, \eqref{CDF2}, and \eqref{CDF3}.
\vspace{-0.3cm}
\begin{cor}\label{cor:1}
Let $X$ and $Y$ be two independent exponential random variables with unity mean, and let $a$, $b$, $I$, and $g$ be given non-negative constants. Then the cdfs of $T_1$ and $T_2$, and the joint cdf of $T_1$ and $T_3$ are given as
\vspace{-0.1cm}
\begin{align}
F_{T_1}(\tau) &= 1 - \frac{a}{a + b\tau} {\rm e}^{-\frac{\tau}{a}I},\label{CDF4}\\
F_{T_2}(\tau) &= 1 - \frac{a}{a + b\tau} {\rm e}^{-\frac{\tau}{a}I} - \frac{b}{b + a\tau} {\rm e}^{-\frac{\tau}{b}I} + \frac{ab(1 + \tau)(1 - \tau)}{(a + b\tau)(b + a\tau)} {\rm e}^{-\left(\frac{1}{a} + \frac{1}{b}\right)\frac{\tau}{|1-\tau|{\bf 1}(\tau < 1)}I},\label{CDF5}\\
F_{T_1, T_3}(\tau, \tau) &= 1 - \frac{a}{a + b\tau} {\rm e}^{-\frac{\tau}{a}I} - \frac{b(1-\tau g)}{b(1-\tau g) + a\tau(1+g)} {\rm e}^{-\frac{\tau(1+g)}{b|1-\tau g|{\bf 1}(\tau < \frac{1}{g})}I}\nonumber\\
&\quad+ \frac{ab(1 + \tau)(1 - \tau(1+g))}{(a + b\tau)(b(1-\tau g) + a\tau(1+g))} {\rm e}^{-\left(\frac{1}{a(1+g)} + \frac{1}{b}\right)\frac{\tau(1+g)}{|1-\tau(1+g)|{\bf 1}(\tau < \frac{1}{1+g})}I}.\label{CDF6}
\end{align}
\end{cor}
\vspace{-0.6cm}
\subsection{Relative Distance and Angle Distributions in the 3D Setting} \label{sec:DistDist}
Let $\tilde{r}_{{\rm B}_0}$, $\tilde{r}_{{\rm D_0, L}}$, and $\tilde{r}_{{\rm D_0, N}}$ be the distances of the closest BS, LoS UAV, and NLoS UAV to the typical UE, and also let $\tilde{\theta}_{{\rm D_0, L}}$ and $\tilde{\theta}_{{\rm D_0, N}}$ be the zenith angles of the closest LoS and NLoS UAVs to the typical UE, respectively. We start by providing the distance distribution of the closest BS to the typical UE in the following lemma, where the proof follows directly from the null probability of a PPP \cite{B_Haenggi_Stochastic_2012} and is omitted here for brevity.
\vspace{-0.3cm}
\begin{lemma}\label{lem:1}
The cdf and pdf of the closest BS distance to the origin, i.e., $\tilde{r}_{{\rm B}_0}$, can be written as
\begin{align} \label{Dist_rTilde_B}
F_{\tilde{r}_{{\rm B}_0}}(r) = 1 - {\rm e}^{-\pi \lambda_{\rm B} (r^2 - h_{\rm B}^2)}, \qquad
f_{\tilde{r}_{{\rm B}_0}}(r) = 2\pi \lambda_{\rm B} r {\rm e}^{-\pi \lambda_{\rm B} (r^2 - h_{\rm B}^2)}.
\end{align}
\end{lemma}
As mentioned earlier in Section \ref{subsec4:AssociationPolicy}, since all BSs are in the NLoS channel condition and have the same height, the closest BS to the typical UE is also regarded as its serving BS. Hence, we have $r_{{\rm B}_0} = \tilde{r}_{{\rm B}_0}$, and thus, the serving BS distance distribution to the typical UE will be $f_{r_{{\rm B}_0}}(r) = f_{\tilde{r}_{{\rm B}_0}}(r)$.
We now provide the relative distance and angle distributions in 3D aerial networks in the following lemmas. First, we derive the distance distribution of the closest UAV to the typical UE for the LoS and NLoS channel conditions in Lemma \ref{lem:2_pre}. Using this result, we obtain the \textit{joint distance and angle} distribution of the closest UAV to the typical UE in Lemma \ref{lem:2}. We will then determine the association probabilities for each channel condition in Lemma \ref{lem:3}. Finally, we derive the joint distribution of the distance and angle of the \textit{serving} UAV to the typical UE for both channel conditions in Lemma \ref{lem:4}.
\vspace{-0.3cm}
\begin{lemma}\label{lem:2_pre}
The cdf and pdf of the closest UAV distance to the origin for channel condition $q = \{\rm L, N\}$, i.e., $\tilde{r}_{{\rm D}_0, q}$, can be written, respectively, as
\begin{align}\label{Dist_rTilde_D}
F_{\tilde{r}_{{\rm D}_0, q}}(r) = 1 - {\rm e}^{-\pi \lambda_{\rm D} \beta_q(r)}, \quad
f_{\tilde{r}_{{\rm D}_0, q}}(r) = 2\pi \lambda_{\rm D} r^2 \left( \int_{\theta_{\rm D, M}(r)}^{\theta_{\rm D, m}(r)}\!\sin(\omega)p_q(\omega)\,{\rm d}\omega \right){\rm e}^{-\pi \lambda_{\rm D} \beta_q(r)},
\end{align}
where $\theta_{\rm D, m}(r) = \cos^{-1}\left(\frac{h_{\rm D, m}}{r}\right)$, $\theta_{\rm D, M}(r) = \cos^{-1}\left(\min\left\{\frac{h_{\rm D, M}}{r}, 1\right\}\right)$, $p_q(\theta)$ is given in \eqref{eq:LoS}, and
\begin{align}
\beta_q(r) &= \frac{2}{3}\int_0^{\theta_{\rm D, m}(r)}\left( \min\{h_{\rm D, M}^3, r^3\cos^3(\theta_1)\} - h_{\rm D, m}^3 \right) \frac{\sin(\theta_1)}{\cos^3(\theta_1)}p_q(\theta_1)\,{\rm d}\theta_1.
\end{align}
\end{lemma}
\begin{IEEEproof}
Consider the 3D setting in Fig. \ref{fig:Setting3D}, where the typical UE is located at the origin and the UAVs are distributed as a 3D homogeneous PPP in the region enclosed between heights $h_{\rm D, m}$ and $h_{\rm D, M}$. Depending on where the closest UAV resides, an \textit{exclusion zone} is formed, where no other UAVs are allowed to enter. This exclusion zone is a spherical cap when $h_{\rm D, m} \leq \tilde{r}_{{\rm D}_0, q} \leq h_{\rm D, M}$ (see Fig. \ref{fig:Setting3D} (a)) and a spherical segment when $\tilde{r}_{{\rm D}_0, q} > h_{\rm D, M}$ (see Fig. \ref{fig:Setting3D} (b)). Therefore, we can write the cdf of $\tilde{r}_{{\rm D}_0, q}$ as
\begin{align*}
F_{\tilde{r}_{{\rm D}_0, q}}(r) &= {\mathbb P}[\tilde{r}_{{\rm D}_0, q} \leq r] = 1 - {\mathbb P}[{\rm No ~ UAV ~ in ~ the ~ spherical ~ cap ~ \mathcal{A} ~ or ~ spherical ~ segment ~ \mathcal{B}}]\\
&= 1 - {\rm e}^{-\Lambda(\mathcal{A})} {\bf 1}(h_{\rm D, m} \leq r \leq h_{\rm D, M}) - {\rm e}^{-\Lambda(\mathcal{B})} {\bf 1}(r > h_{\rm D, M})\\
&\overset{(a)}{=} 1 - \begin{cases}
\exp\Big[ -\int_0^{2\pi}\int_0^{\cos^{-1}(\frac{h_{\rm D, m}}{r})}\int_{\frac{h_{\rm D, m}}{\cos(\theta_1)}}^{r} \lambda_{\rm D}p_q(\theta_1) r_1^2\sin(\theta_1)\, {\rm d}r_1\, {\rm d}\theta_1\, {\rm d}\phi_1 \Big] & r \leq h_{\rm D, M}\\
\exp\Big[ -\int_0^{2\pi}\int_0^{\cos^{-1}(\frac{h_{\rm D, M}}{r})}\int_{\frac{h_{\rm D, m}}{\cos(\theta_1)}}^{\frac{h_{\rm D, M}}{\cos(\theta_1)}} \lambda_{\rm D}p_q(\theta_1) r_1^2\sin(\theta_1)\, {\rm d}r_1\, {\rm d}\theta_1\, {\rm d}\phi_1\\
\hspace{1cm}-\int_0^{2\pi}\int_{\cos^{-1}(\frac{h_{\rm D, M}}{r})}^{\cos^{-1}(\frac{h_{\rm D, m}}{r})}\int_{\frac{h_{\rm D, m}}{\cos(\theta_1)}}^{r} \lambda_{\rm D}p_q(\theta_1) r_1^2\sin(\theta_1)\, {\rm d}r_1\, {\rm d}\theta_1\, {\rm d}\phi_1 \Big] & r > h_{\rm D, M}
\end{cases},
\end{align*}
where $\Lambda(\ncalS)$ is the intensity measure of set $\ncalS$ and $(r_1, \theta_1, \phi_1)$ is the spherical coordinate triplet. Note that in $(a)$ we derived the null probability of 3D PPP $\Phi_{{\rm D}, q}$ by integrating its density (i.e., $\lambda_{\rm D}p_q(\theta_1)$) over the spherical cap $\mathcal{A}$ for $h_{\rm D, m} \leq r \leq h_{\rm D, M}$ and the spherical segment $\mathcal{B}$ for $r > h_{\rm D, M}$. Evaluating these integrals and taking their derivatives with respect to $r$, we end up with the cdf and pdf of $\tilde{r}_{{\rm D}_0, q}$ as given in \eqref{Dist_rTilde_D}.
\begin{figure}[!t]
\centering
\includegraphics[width=0.8\columnwidth]{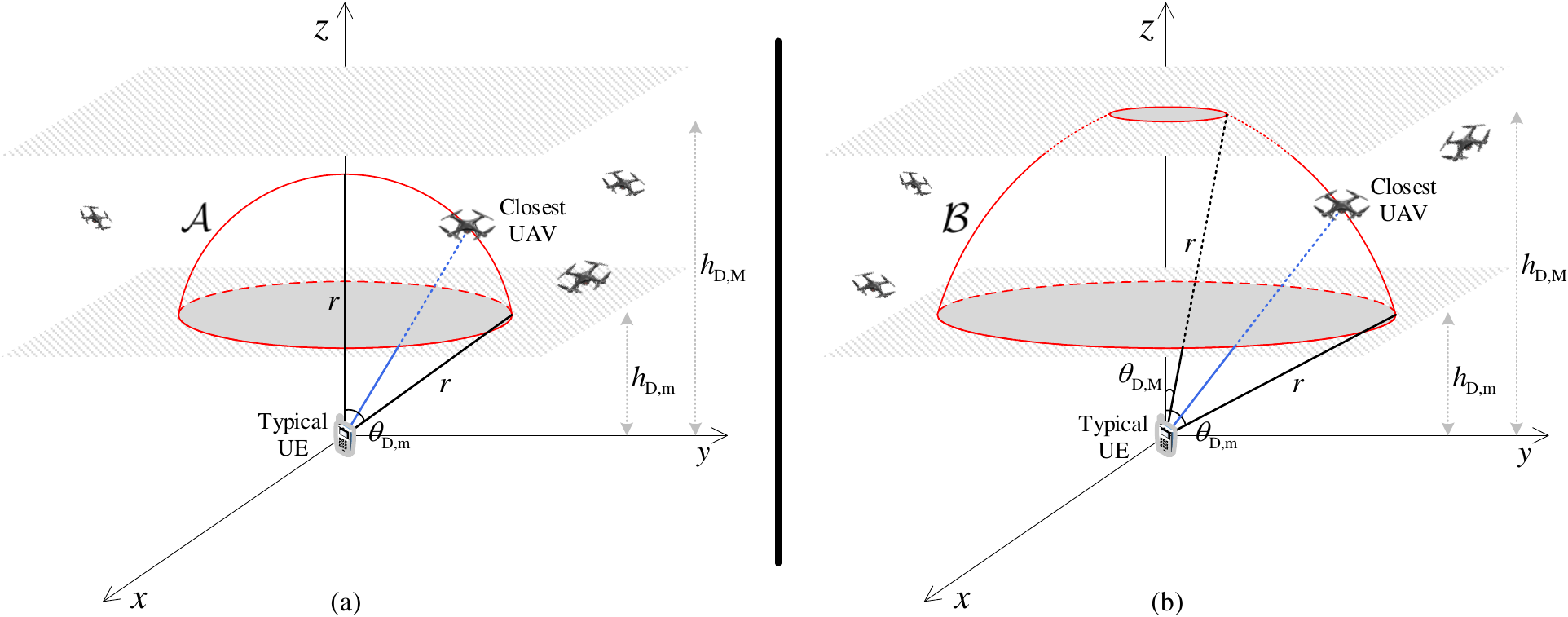}
\vspace{-0.6cm}
\caption{The 3D network setting when the closest UAV is at distance $r$ from $\nbo$ for (a) $h_{\rm D, m} \leq r \leq h_{\rm D, M}$, and (b) $r > h_{\rm D, M}$.}
\vspace{-0.9cm}
\label{fig:Setting3D}
\end{figure}
\end{IEEEproof}
\vspace{-0.3cm}
\begin{lemma}\label{lem:2}
The joint pdf of the distance and zenith angle of the closest UAV to the origin for channel condition $q$, i.e., $\tilde{r}_{{\rm D}_0, q}$ and $\tilde{\theta}_{{\rm D}_0, q}$, respectively, can be written as
\begin{align} \label{Dist_rTilde_D_pdf}
f_{\tilde{r}_{{\rm D}_0, q}, \tilde{\theta}_{{\rm D}_0, q}}\!(r, \theta) \!=\! 2\pi \lambda_{\rm D} r^2\! \left( \int_{\theta_{\rm D, M}(r)}^{\theta_{\rm D, m}(r)}\!\!\!\scalebox{0.95}{$\sin(\omega)p_q(\omega)\,{\rm d}\omega$} \!\right)\! \frac{\scalebox{0.93}{${\rm e}^{-\pi \lambda_{\rm D} \beta_q(r)}\sin(\theta){\bf 1}(\theta_{\rm D, M}(r) \!\leq\! \theta \!\leq\! \theta_{\rm D, m}(r))$}}{\scalebox{0.97}{$\cos(\theta_{\rm D, M}(r)) - \cos(\theta_{\rm D, m}(r))$}},
\end{align}
where $\theta_{\rm D, m}(r)$, $\theta_{\rm D, M}(r)$, and $\beta_q(r)$ are as given in Lemma \ref{lem:2_pre}.
\end{lemma}
\vspace{-0.3cm}
\begin{IEEEproof}
To obtain the joint pdf of $\tilde{r}_{{\rm D}_0, q}$ and $\tilde{\theta}_{{\rm D}_0, q}$, we first derive the conditional pdf $f_{\tilde{\theta}_{{\rm D}_0, q}|\tilde{r}_{{\rm D}_0, q}}(\theta|r)$. Conditioned on $\tilde{r}_{{\rm D}_0, q} = r$, the closest UAV is distributed uniformly on the surface of the spherical cap $\ncalA$ or the spherical segment $\ncalB$. Since the differential element of solid angle $\Omega$ for a sphere is given by ${\rm d}\Omega = r^2\sin(\theta){\rm d}r{\rm d}\theta{\rm d}\phi = -r^2{\rm d}r{\rm d}(\cos(\theta)){\rm d}\phi$, we conclude that $\cos(\theta)$ should be uniformly distributed between $\cos(\theta_{\rm D, M}(r))$ and $\cos(\theta_{\rm D, m}(r))$. Hence, we have
\begin{align*}
f_{\tilde{\theta}_{{\rm D}_0, q}|\tilde{r}_{{\rm D}_0, q}}(\theta|r) = \frac{\sin(\theta)}{\cos(\theta_{\rm D, M}(r)) - \cos(\theta_{\rm D, m}(r))} {\bf 1}(\theta_{\rm D, M}(r) \!\leq\! \theta \!\leq\! \theta_{\rm D, m}(r)).
\end{align*}
Now, using \scalebox{0.92}{$f_{\tilde{r}_{{\rm D}_0, q}, \tilde{\theta}_{{\rm D}_0, q}}(r, \theta) = f_{\tilde{\theta}_{{\rm D}_0, q}|\tilde{r}_{{\rm D}_0, q}}(\theta|r) f_{\tilde{r}_{{\rm D}_0, q}}(r)$} and the previous lemma, we arrive at \eqref{Dist_rTilde_D_pdf}.
\end{IEEEproof}
The random variables $\tilde{r}_{{\rm D}_0, q}$ and $\tilde{\theta}_{{\rm D}_0, q}$ are clearly dependent. However, this dependency becomes less significant as $h_{\rm D, m} \to 0$ and $h_{\rm D, M} \to \infty$. In fact, when UAVs are distributed as a PPP in the half-space $z \geq 0$, $\tilde{r}_{{\rm D}_0, q}$ and $\tilde{\theta}_{{\rm D}_0, q}$ are independent from each other and we have
\begin{align}
f_{\tilde{r}_{{\rm D}_0, q}}(r) = 2\pi \lambda_{\rm D} b_q r^2 \, {\rm e}^{-\frac{2}{3}\pi \lambda_{\rm D} b_q r^3}, \qquad f_{\tilde{\theta}_{{\rm D}_0, q}}(\theta) = \sin(\theta){\bf 1}\left(0 \leq \theta \leq \frac{\pi}{2}\right),\label{Dist_rTilde_thetaTilde_pdf}
\end{align}
where $b_q = \int_{0}^{\frac{\pi}{2}}\sin(\omega)p_q(\omega)\,{\rm d}\omega$
\begin{lemma}\label{lem:3}
The probability that the typical UE is associated with an NLoS UAV is given as
\begin{align}\label{Association_L_D}
A_{\rm N} = \int_{h_{\rm D, m}}^{\infty} \int_{\theta_{\rm D, M}(r)}^{\theta_{\rm D, m}(r)} 2\pi \lambda_{\rm D} r^2 \sin(\theta) p_{\rm N}(\theta) {\rm e}^{-\pi \lambda_{\rm D} \left(\beta_{\rm N}(r) + \beta_{\rm L}\left( \left(\frac{\eta_{\rm N}}{\eta_{\rm L}}\right)^{\frac{1}{\alpha_{\rm L}}} r^{\frac{\alpha_{\rm N}}{\alpha_{\rm L}}} \right) \right)}\,{\rm d}\theta\,{\rm d}r
\end{align}
where $\theta_{\rm D, m}(r)$, $\theta_{\rm D, M}(r)$, $\beta_{\rm N}(r)$, and $\beta_{\rm L}(r)$ are as given in Lemma \ref{lem:2_pre}. Furthermore, the probability that the typical UE is associated with an LoS UAV is $A_{\rm L} = 1 - A_{\rm N}$.
\end{lemma}
\vspace{-0.3cm}
\begin{IEEEproof}
See Appendix \ref{app:lem:3}.
\end{IEEEproof}
\vspace{-0.3cm}
\begin{lemma}\label{lem:4}
Given that the typical UE is associated with UAV ${\rm D_0}$ with channel condition $q$, the joint pdf of the serving distance and angle between ${\rm D_0}$ and the typical UE can be written as 
\begin{align}\label{Serving_L_D}
f_{r_{{\rm D_0}, q}, \theta_{{\rm D_0}, q}}(r, \theta) = \frac{1}{A_q} f_{\tilde{r}_{{\rm D_0}, q}, \tilde{\theta}_{{\rm D_0}, q}}(r, \theta) \exp\left[-\pi \lambda_{\rm D} \beta_{\bar{q}}\left(\max\left\{ h_{\rm D, m}, \left(\frac{\eta_q}{\eta_{\bar{q}}}\right)^{\frac{1}{\alpha_{\bar{q}}}} r^{\frac{\alpha_q}{\alpha_{\bar{q}}}} \right\} \right)\right],
\end{align}
where $\bar{q} = \{{\rm L, N}\}\setminus q$ and $f_{\tilde{r}_{\rm D_0, L}, \tilde{\theta}_{\rm D_0, L}}(r)$ and $\beta_q(r)$ are given in Lemmas \ref{lem:2} and \ref{lem:2_pre}, respectively.
\end{lemma}
\vspace{-0.3cm}
\begin{IEEEproof}
See Appendix \ref{app:lem:4}.
\end{IEEEproof}
\vspace{-0.4cm}
\section{Performance Analysis} \label{sec:PerformanceEvaluation}
In this section, we will first derive the conditional Laplace transform of the interference imposed at the typical UE by both the BSs and the UAVs. Using this result along with the results of the previous section, we will be ready to analyze the coverage probability.
\vspace{-0.4cm}
\subsection{Conditional Laplace Transform of Interference} \label{sec:LaplaceTransformInterference}
The Laplace transform of random variable $X$ at point $s$ is defined as $\ncalL_X(s) = \nbbE[{\rm e}^{-sX}]$. Assuming $X = I_{\rm U}$, i.e., the total interference at the typical UE, our goal is to derive the Laplace transform of $I_{\rm U}$ conditioned on knowing the locations of the serving BS and UAV, i.e., ${\rm B_0}$ and ${\rm D_0}$, respectively. Since the channel condition of the serving UAV affects the total interference, we further condition the Laplace transform on the serving UAV being in LoS or NLoS conditions. Since $I_{\rm BU}$, $I_{\rm DU, L}$, and $I_{\rm DU, N}$ are independent from each other given ${\rm B_0}$ and ${\rm D_0}$, we have
\begin{align}\label{Lap_I_U}
\ncalL_{I_{{\rm U}|q}}(s|{\rm B_0}, {\rm D_0}) = \ncalL_{I_{\rm BU}}(s|{\rm B_0}) \ncalL_{I_{{\rm DU, L}|q}}(s|{\rm D_0}) \ncalL_{I_{{\rm DU, N}|q}}(s|{\rm D_0}),
\end{align}
where $q = \{\rm L, N\}$ represents the serving UAV channel condition.

As mentioned in Section \ref{subsubsec3.1:ReceivedPowers}, $\Phi_{\rm B}'$ is an inhomogeneous PPP with density $\lambda_{\rm B}$ for $u_{{\rm B}_\nbx} \geq u_{{\rm B}_0}$ and $0$ otherwise. Since all BSs experience the NLoS channel condition, there is an exclusion zone $\ncalX_{\rm B} = b(\nbo, u_{{\rm B}_0})$ for the projection of interfering BSs onto the ground, where $b(\nbo, r)$ is a disc of radius $r$ centered at $\nbo$. Next, we derive the conditional Laplace transform of $I_{\rm BU}$.
\begin{table}[!t]
	\centering
	\caption{\label{table:ExcZone}Exclusion zone radius for different interfering and serving UAV channel conditions.}
	\vspace{-0.3cm}
	\begin{tabular}{|c|c|c||c|c|c|}
		\hline
		Interfering UAVs & Serving UAV & Exclusion zone radius & Interfering UAVs & Serving UAV & Exclusion zone radius \\
		\hline\hline
		LoS & LoS & $r_{\rm L|L} = r$ & NLoS & LoS & $r_{\rm N|L} = \left(\frac{\eta_{\rm L}}{\eta_{\rm N}}\right)^{\frac{1}{\alpha_{\rm N}}} r^{\frac{\alpha_{\rm L}}{\alpha_{\rm N}}}$ \\ 
		\hline
		LoS & NLoS & $r_{\rm L|N} = \left(\frac{\eta_{\rm N}}{\eta_{\rm L}}\right)^{\frac{1}{\alpha_{\rm L}}} r^{\frac{\alpha_{\rm N}}{\alpha_{\rm L}}}$ & NLoS & NLoS & $r_{\rm N|N} = r$ \\
		\hline
	\end{tabular}
	\vspace{-0.8cm}
\end{table}
%
%
%
\vspace{-0.3cm}
\begin{lemma}\label{lem:5}
The Laplace transform of interference from the BSs at the typical UE conditioned on knowing the location of the serving BS (with 2D distance $u_{\rm B_0}$ to $\nbo$) can be written as
\begin{align}\label{Lap_I_BU}	
\ncalL_{I_{\rm BU}}\!(s|{\rm B_0}) \!=\! \exp\!\left[ -2\pi\lambda_{\rm B}\!\!\int_{u_{\rm B_0}}^{\infty} \!\!\left[ 1 \!- \!\!\left( \!1 \!+\! \frac{sP_{\rm B}}{m\eta_{\rm N}}\frac{G_{\rm B}^{\rm OmniD}\!\left(\pi \!-\! \tan^{-1}(\frac{u_{{\rm B}_\nbx}}{h_{\rm B}}), \theta_{\rm B}\right)}{\left(u_{{\rm B}_\nbx}^2 \!+ h_{\rm B}^2\right)^\frac{\alpha_{\rm N}}{2}} \right)^{\!\!\!\!\!-m} \right] \!\!u_{{\rm B}_\nbx} {\rm d}u_{{\rm B}_\nbx} \right]\!\!,
\end{align}
where $G_{\rm B}^{\rm OmniD}(\theta, \theta_{\rm B})$ is the downtilted BS antenna gain along direction $\theta$, which is given in \eqref{G_OmniD}.
\end{lemma}
\vspace{-0.3cm}
\begin{IEEEproof}
See Appendix \ref{app:lem:5}.
\end{IEEEproof}
Since UAVs experience both LoS and NLoS channel conditions, the exclusion zone for the interfering UAVs, i.e., $\ncalX_{\rm D}$, depends on the channel conditions of both the serving UAV and the interfering UAVs. Assuming that the serving UAV has distance $r$ to the origin, $\ncalX_{\rm D}$ will either be a spherical cap or a spherical segment with radius $r_{q_1|q_2}$, where $q_1, q_2=\{\rm L, N\}$ denote the channel condition of the interfering UAVs and the serving UAV, respectively. Note that $\ncalX_{\rm D}$ is a spherical cap when $h_{\rm D, m} \leq r_{q_1|q_2} \leq h_{\rm D, M}$ and a spherical segment when $r_{q_1|q_2} > h_{\rm D, M}$. The exclusion zone radii for different values of $q_1$ and $q_2$ are given in Table \ref{table:ExcZone}, using which we derive the conditional Laplace transform of $I_{{\rm DU}, q_1|q_2}$ in the next lemma.
\vspace{-0.3cm}
\begin{lemma}\label{lem:6}
The Laplace transform of interference from the UAVs with channel condition $q_1$ at the typical UE conditioned on knowing the location and channel condition of the serving UAV (with 3D distance $r_{{\rm D_0},q_2}$ to $\nbo$, where $q_2$ is the serving UAV channel condition) can be written as
\begin{align}\label{Lap_I_DU}	
\ncalL_{I_{{\rm DU}, q_1|q_2}}\!(s|{\rm D_0}) \!&= \!\exp\!\Bigg[\! -\!2\pi\lambda_{\rm D}\!\int_{\theta_{\rm D, M}}^{\theta_{\rm D, m}}\!\! \!\int_{r_{q_1|q_2}}^{\frac{h_{\rm D, M}}{\cos(\theta)}} \!\bigg[ 1 \!- \!\Big( 1 \!+\! \frac{sP_{\rm D}}{m\eta_{q_1}}\frac{G_{\rm D}^{\rm AC}\!\left(\pi \!- \theta\right)}{r^{\alpha_{q_1}}} \Big)^{\!\!-m} \bigg] p_{q_1}(\theta) r^2\!\sin(\theta) {\rm d}r {\rm d}\theta \nonumber\\
&-2\pi\lambda_{\rm D}\!\int_{\theta_{\rm D, m}}^{\frac{\pi}{2}}\! \int_{\frac{h_{\rm D, m}}{\cos(\theta)}}^{\frac{h_{\rm D, M}}{\cos(\theta)}} \!\bigg[ 1 \!- \!\Big( 1 \!+\! \frac{sP_{\rm D}}{m\eta_{q_1}}\frac{G_{\rm D}^{\rm AC}\!\left(\pi \!- \theta\right)}{r^{\alpha_{q_1}}} \Big)^{\!\!-m} \bigg] p_{q_1}(\theta) r^2\!\sin(\theta) {\rm d}r {\rm d}\theta \Bigg],
\end{align}
where $\theta_{\rm D, m} = \cos^{-1}(\frac{h_{\rm D, m}}{r_{q_1|q_2}})$, $\theta_{\rm D, M} = \cos^{-1}\left(\min\left\{\frac{h_{\rm D, M}}{r_{q_1|q_2}}, 1\right\}\right)$, and $G_{\rm D}^{\rm AC}\left(\theta\right)$ is the UAV access antenna gain along direction $\theta$, which is given in dBi in \eqref{G_UAV}.
\end{lemma}
\vspace{-0.3cm}
\begin{IEEEproof}
See Appendix \ref{app:lem:6}.
\end{IEEEproof}
Using the results of Lemmas \ref{lem:5} and \ref{lem:6}, we end up with the conditional Laplace transform of interference at the typical UE given the serving UAV has channel condition $q$ as in \eqref{Lap_I_U}.
\vspace{-0.5cm}
\subsection{Coverage Probability} \label{sec:CoverageProbability}
In this section, we derive the coverage probability for the typical UE considering both AF and DF relaying protocols. Since we assumed a hybrid scheme in this paper, the received SINR at the typical UE is the maximum of the BS-UE SINR and the relay-aided end-to-end SINR, as given in \eqref{eq:SINR}. The following two theorems provide the main results of this paper.
\vspace{-0.3cm}
\begin{theorem} \label{theorem1}
The network coverage probability for the AF protocol can be written as $P_{\rm Cov}^{\rm AF} \!=\! A_{\rm L}P_{\rm Cov, L}^{\rm AF} \!+\! A_{\rm N}P_{\rm Cov, N}^{\rm AF}$,
where $A_q$ is the probability that the typical UE is associated with a UAV with channel condition $q$ as given in Lemma \ref{lem:3}, and $P_{{\rm Cov}, q}^{\rm AF}$ is the ccdf of ${\rm SINR}_q^{\rm AF}$, given as
\begin{align}\label{eq2:thm1}
P_{{\rm Cov}, q}^{\rm AF} \!=\! \sum_{i=1}^3\! \underset{\ncalR \cap \hat{\ncalR}_i}{\int \!\!\dots\! \int} \scalebox{0.865}{$\!W_i f_Z(z) f_{r_{{\rm B}_0}}(r_{{\rm B}_0}) f_{r_{{\rm D}_0, q}, \theta_{{\rm D}_0, q}}(r_{{\rm D}_0, q}, \theta_{{\rm D}_0, q}) f_{\phi_{\rm B_0 D_0}}(\phi_{\rm B_0 D_0}) {\rm d}z{\rm d}r_{{\rm B}_0}{\rm d}\theta_{{\rm D}_0, q}{\rm d}r_{{\rm D}_0, q}{\rm d}\phi_{\rm B_0 D_0}$},
\end{align}
where $Z=f_{\rm B_0 D_0} \sim {\rm Gamma}(m, m)$, $f_{r_{{\rm B}_0}}(r)$ and $f_{r_{{\rm D}_0, q}, \theta_{{\rm D}_0, q}}(r, \theta)$ are given in Lemmas \ref{lem:1} and \ref{lem:4}, respectively, $\phi_{\rm B_0 D_0} \sim U[0, 2\pi)$ is the azimuthal angle between ${\rm B_0}$ and ${\rm D_0}$, $\ncalR = \{0 \leq z < \infty, h_{\rm B} \leq r_{{\rm B}_0} < \infty, \theta_{\rm D, M}(r) \leq \theta_{{\rm D}_0, q} \leq \theta_{\rm D, m}(r), h_{\rm D, m} \leq r_{{\rm D}_0, q} < \infty, 0 \leq \phi_{\rm B_0 D_0} < 2\pi\}$, $\hat{\ncalR}_1 = \{\frac{1}{g} \leq \tau\}$, $\hat{\ncalR}_2 = \{\frac{1}{1 + g} \leq \tau < \frac{1}{g}\}$, $\hat{\ncalR}_3 = \{\tau < \frac{1}{1+g}\}$, and $W_i$'s are defined as follows:
\begin{align*}
&W_1 = \sum_{i=0}^{m-1}\sum_{k=0}^{i} {k+m-1 \choose k} \mu\bigg(\begin{matrix} a,  b\tau \\ m,  k\end{matrix} \,\bigg|\, \begin{matrix} s_1,  s_1 \\ i-k \end{matrix} \bigg), \\
&W_2 = W_1 + \sum_{i=0}^{m-1}\sum_{k=0}^{i} {k+m-1 \choose k} \mu\bigg(\begin{matrix} a\tau (1+g),  b(1-\tau g) \\ k,  m\end{matrix} \,\bigg|\, \begin{matrix} s_2,  s_2 \\ i-k \end{matrix} \bigg), \\
&W_3 = W_2 + \sum_{i=0}^{m-1} \sum_{j=0}^{i+m-1} {i+m-1 \choose i} \left[\mu\bigg(\begin{matrix} a(1+g),  b \\ m, i\end{matrix} \,\bigg|\, \begin{matrix} s_3,  s_3 \\ j \end{matrix} \bigg) + \mu\bigg(\begin{matrix} a(1+g),  b \\ i, m\end{matrix} \,\bigg|\, \begin{matrix} s_3,  s_3 \\ j \end{matrix} \bigg)\right] \\
&\qquad - \sum_{i=0}^{m-1}\sum_{k=0}^{i}\sum_{j=0}^{k+m-1} {k+m-1 \choose k} {j+i-k \choose j} \Bigg[\left(\frac{\tau(1+g)}{1-\tau(1+g)}\right)^j \mu\bigg(\begin{matrix} a,  b\tau \\ m,  k-j\end{matrix} \,\bigg|\, \begin{matrix} s_1,  s_3 \\ j+i-k \end{matrix} \bigg)\\
&\quad\qquad + \left(\frac{\tau}{1-\tau(1+g)}\right)^j  \mu\bigg(\begin{matrix} a\tau (1+g),  b(1-\tau g) \\ k-j,  m\end{matrix} \,\bigg|\, \begin{matrix} s_2,  s_3 \\ j+i-k \end{matrix} \bigg)\Bigg],\\
&\mu\bigg(\begin{matrix} x, y \\ i, j\end{matrix} \,\bigg|\, \begin{matrix} r, s \\ k \end{matrix} \bigg) = \frac{x^i y^j}{(x + y)^{i + j}} \frac{\left(-r\right)^{k}}{k!} \frac{\partial^{k}}{\partial s^{k}}{\rm e}^{-sN_0} \ncalL_{I_{{\rm U}|q}}(s|{\rm B_0}, {\rm D_0}),\\
&s_1 = \frac{1}{a}m\tau, \qquad s_2 = \frac{1+g}{b(1-\tau g)}m\tau, \qquad s_3 = \frac{a(1+g) + b}{ab(1-\tau(1+g))}m\tau,\\
&a = P_{\rm B} G_{{\rm B}_0} r_{{\rm B}_0}^{-\alpha_{\rm N}} \eta_{\rm N}^{-1}, \qquad b = P_{\rm D} G_{{\rm D}_0} r_{{\rm D_0}, q}^{-\alpha_q} \eta_q^{-1}, \qquad c = P_{\rm B} g_{{\rm B}_0} g_{{\rm D}_0} r_{{\rm B}_0{\rm D}_0}^{-\alpha_{\rm L}} \eta_{\rm L}^{-1}, \qquad g = \frac{N_0}{c Z},\\
&r_{{\rm B}_0{\rm D}_0}^2 = r_{{\rm B}_0}^2 + r_{{\rm D_0}, q}^2 - 2h_{\rm B}r_{{\rm D_0}, q}\cos(\theta_{{\rm D_0}, q}) - 2\sqrt{r_{{\rm B_0}}^2 - h_{\rm B}^2}\, r_{{\rm D_0}, q} \sin(\theta_{{\rm D_0}, q})\cos(\phi_{{\rm B}_0{\rm D}_0}),
\end{align*}
where $G_{{\rm B}_0} = G_{\rm B}^{\rm OmniD}(\pi - \cos^{-1}(\frac{h_{\rm B}}{r_{{\rm B}_0}}), \theta_{\rm B})$, $G_{{\rm D}_0} = G_{\rm D}^{\rm AC}(\pi - \theta_{{\rm D_0}, q})$, and $g_{{\rm D}_0} = G^{\max}$. As for the BS backhaul antenna gain, we have $g_{{\rm B}_0} = G_{\rm B}^{\rm OmniD}(\cos^{-1}(\frac{r_{{\rm D_0}, q}\cos(\theta_{{\rm D_0}, q}) - h_{\rm B}}{r_{{\rm B}_0{\rm D}_0}}), \theta_{\rm B})$ and $g_{{\rm B}_0} = G^{\max}$ for the BS first (Section \ref{par:OmniBS}) and second (Section \ref{par:OmniDirBS}) antenna models, respectively.
\end{theorem}
\vspace{-0.3cm}
\begin{IEEEproof}
See Appendix \ref{app:theorem1}.
\end{IEEEproof}
\vspace{-0.3cm}
\begin{theorem}\label{theorem2}
The network coverage probability for the DF protocol can be written as $P_{\rm Cov}^{\rm DF} \!=\! A_{\rm L}P_{\rm Cov, L}^{\rm DF} \!+\! A_{\rm N}P_{\rm Cov, N}^{\rm DF}$,
where $A_q$ is given in Lemma \ref{lem:3} and $P_{{\rm Cov}, q}^{\rm DF}$ is the ccdf of ${\rm SINR}_q^{\rm DF}$, given as
\begin{align}\label{eq2:thm2}
P_{{\rm Cov}, q}^{\rm DF} = \underset{\ncalR}{\int \!\!\dots\! \int} \scalebox{0.9}{$W f_Z(z) f_{r_{{\rm B}_0}}(r_{{\rm B}_0}) f_{r_{{\rm D}_0, q}, \theta_{{\rm D}_0, q}}(r_{{\rm D}_0, q}, \theta_{{\rm D}_0, q}) f_{\phi_{\rm B_0 D_0}}(\phi_{\rm B_0 D_0}) {\rm d}z{\rm d}r_{{\rm B}_0}{\rm d}\theta_{{\rm D}_0, q}{\rm d}r_{{\rm D}_0, q}{\rm d}\phi_{\rm B_0 D_0}$},
\end{align}
where the region $\ncalR$ and the joint distribution of $Z$, $r_{{\rm B}_0}$, $\theta_{{\rm D}_0, q}$, $r_{{\rm D}_0, q}$, and $\phi_{\rm B_0 D_0}$ are the same as given in Theorem \ref{theorem1}, and $W = V_1 + (1 - V_0)(V_2 + V_3 {\bf 1}(\tau < 1))$, where $V_0 = \frac{\gamma\left( m, \frac{N_0}{c}m\tau \right)}{(m-1)!}$ and 
\begin{align*}
V_1 &= \sum_{i=0}^{m-1}\sum_{k=0}^{i} {k+m-1 \choose k} \mu\bigg(\begin{matrix} a, b\tau \\ m, k\end{matrix} \,\bigg|\, \begin{matrix} s_1, s_1 \\ i-k \end{matrix} \bigg), \\
V_2 &= \sum_{i=0}^{m-1}\sum_{k=0}^{i} {k+m-1 \choose k} \mu\bigg(\begin{matrix} a\tau, b \\ k, m\end{matrix} \,\bigg|\, \begin{matrix} s_2, s_2 \\ i-k \end{matrix} \bigg), \\
V_3 &= \sum_{i=0}^{m-1} \sum_{j=0}^{i+m-1} {i+m-1 \choose i}\left[ \mu\bigg(\begin{matrix} a, b \\ m, i\end{matrix} \,\bigg|\, \begin{matrix} s_3, s_3 \\ j \end{matrix} \bigg) + \mu\bigg(\begin{matrix} a, b \\ i, m\end{matrix} \,\bigg|\, \begin{matrix} s_3, s_3 \\ j \end{matrix} \bigg) \right]\\
&\quad - \sum_{i=0}^{m-1}\sum_{k=0}^{i}\sum_{j=0}^{k+m-1} {k+m-1 \choose k} {j+i-k \choose j} \left(\frac{\tau}{1-\tau}\right)^j \times\\
&\quad\qquad\left[\mu\bigg(\begin{matrix} a, b\tau \\ m, k-j\end{matrix} \,\bigg|\, \begin{matrix} s_1, s_3 \\ j+i-k \end{matrix} \bigg) + \mu\bigg(\begin{matrix} a\tau, b \\ k-j, m\end{matrix} \,\bigg|\, \begin{matrix} s_2, s_3 \\ j+i-k \end{matrix} \bigg) \right],\\
s_1 &= \frac{1}{a}m\tau, \qquad s_2 = \frac{1}{b}m\tau, \qquad s_3 = \frac{a + b}{ab(1-\tau)}m\tau,
\end{align*}
and the function $\mu$ and parameters $a$, $b$, and $c$ are the same as given in Theorem \ref{theorem1}.
\end{theorem}
\vspace{-0.3cm}
\begin{IEEEproof}
See Appendix \ref{app:theorem2}.
\end{IEEEproof}
Ignoring noise and using Assumption \ref{Assumption1}, we have ${\rm SINR_{BD}} \to \infty$, and thus ${{\rm SINR}_{{\rm e2e}, q}^{\rm AF}} = {{\rm SINR}_{{\rm e2e}, q}^{\rm DF}} = {{\rm SINR}_{{\rm DU}, q}}$. The following corollary gives the coverage probability in this scenario, where the proof follows by setting $N_0 = 0$ in Theorems \ref{theorem1} and \ref{theorem2}, giving $g = 0$ and $V_0 = 0$.
\vspace{-0.3cm}
\begin{cor}\label{cor3}
The coverage probability for both the AF and DF protocols in an interference-limited network can be written as $P_{\rm Cov}\!=\! A_{\rm L}P_{\rm Cov, L} \!+\! A_{\rm N}P_{\rm Cov, N}$, where $P_{{\rm Cov}, q}$ is as given in \eqref{eq2:thm2} with $W = V_1 + V_2 + V_3 {\bf 1}(\tau < 1)$, where $V_1$, $V_2$, and $V_3$ are given in Theorem \ref{theorem2} statement.
\end{cor}
\vspace{-0.6cm}
\section{Simulation Results} \label{sec:ITbounds}
In this section, we verify our analytical results via numerical simulations and provide several system-level insights of our 3D setup. We assume that the density of BSs is $\lambda_{\rm B} = 10^{-6}$ (i.e., $1~{\rm BS/km^2}$) and they are located at a constant height of $h_{\rm B} = 20$ m. The BSs provide wireless backhaul connections for the UAVs, which are distributed as a 3D PPP with density $\lambda_{\rm D}$ between heights $h_{\rm D, m}$ and $h_{\rm D, M}$, where we assume $\lambda_{\rm D}$ ranges from $10^{-9}$ to $10^{-6}$, and $h_{\rm D, m}$ and $h_{\rm D, M}$ take values in $\{50, \dots, 1000\}$ m. Following \cite{J_AlHourani_Optimal_2014}, we consider the following four urban environments, where the parameters of the LoS probability function ($c_1$ and $c_2$) and the mean excessive path-loss ($\eta_{\rm L}$ and $\eta_{\rm N}$) for each environment are also provided: (i) suburban ($c_1 = 4.88$, $c_2 = 0.43$, $\eta_{\rm L} = 0.1$ dB, $\eta_{\rm N} = 21$ dB), (ii) urban ($c_1 = 9.61$, $c_2 = 0.16$, $\eta_{\rm L} = 1$ dB, $\eta_{\rm N} = 20$ dB), (iii) dense urban ($c_1 = 12.08$, $c_2 = 0.11$, $\eta_{\rm L} = 1.6$ dB, $\eta_{\rm N} = 23$ dB), and (iv) highrise urban ($c_1 = 27.23$, $c_2 = 0.08$, $\eta_{\rm L} = 2.3$ dB, $\eta_{\rm N} = 34$ dB). Other parameters are $\alpha_{\rm L} = 2.5$, $\alpha_{\rm N} = 4$, $m = \{1, 2\}$, $P_{\rm B} = 10$ dB, $P_{\rm D} = 5$ dB, $N_0 = 10^{-8}$, $N_{\rm B} = 8$, and $\theta_{\rm B} = 100^{\circ}$ (measured from the $z$-axis). 

\begin{figure}[t!]
\centering
\begin{minipage}{0.48\columnwidth}
\centering
\includegraphics[width=0.8\textwidth]{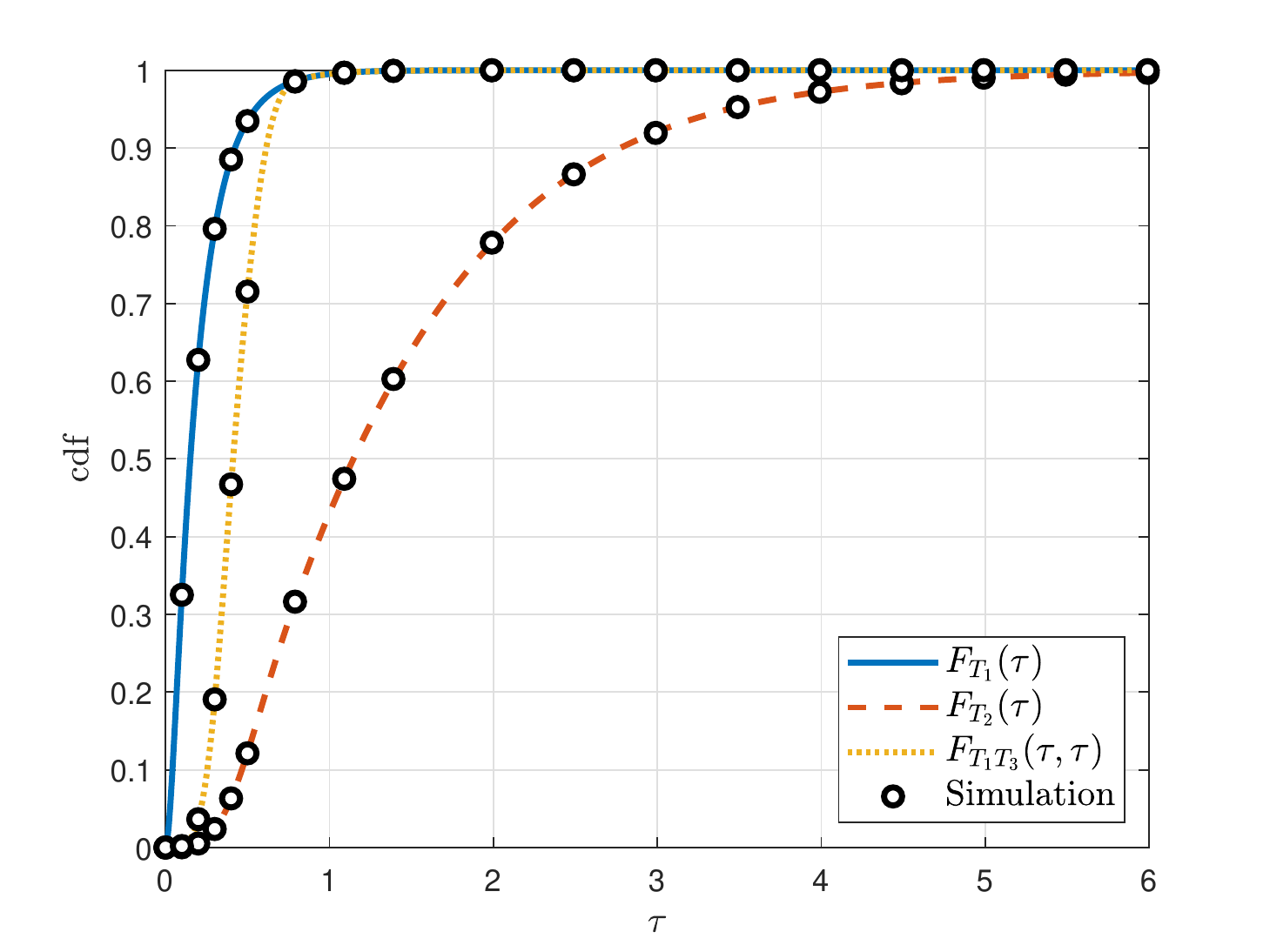}
\vspace{-0.7cm}
\caption{Comparison between the cdfs of $T_1$, $T_2$, and the joint cdf of $T_1$ and $T_3$ ($a = 1$, $b = 4$, $I = 2$, $m = 2$, and $g = 1$).}
\vspace{-0.5cm}
\label{fig:CDF_T1T2T3_All}
\end{minipage}\hfill
\begin{minipage}{0.48\columnwidth}
\centering
\includegraphics[width=0.8\textwidth]{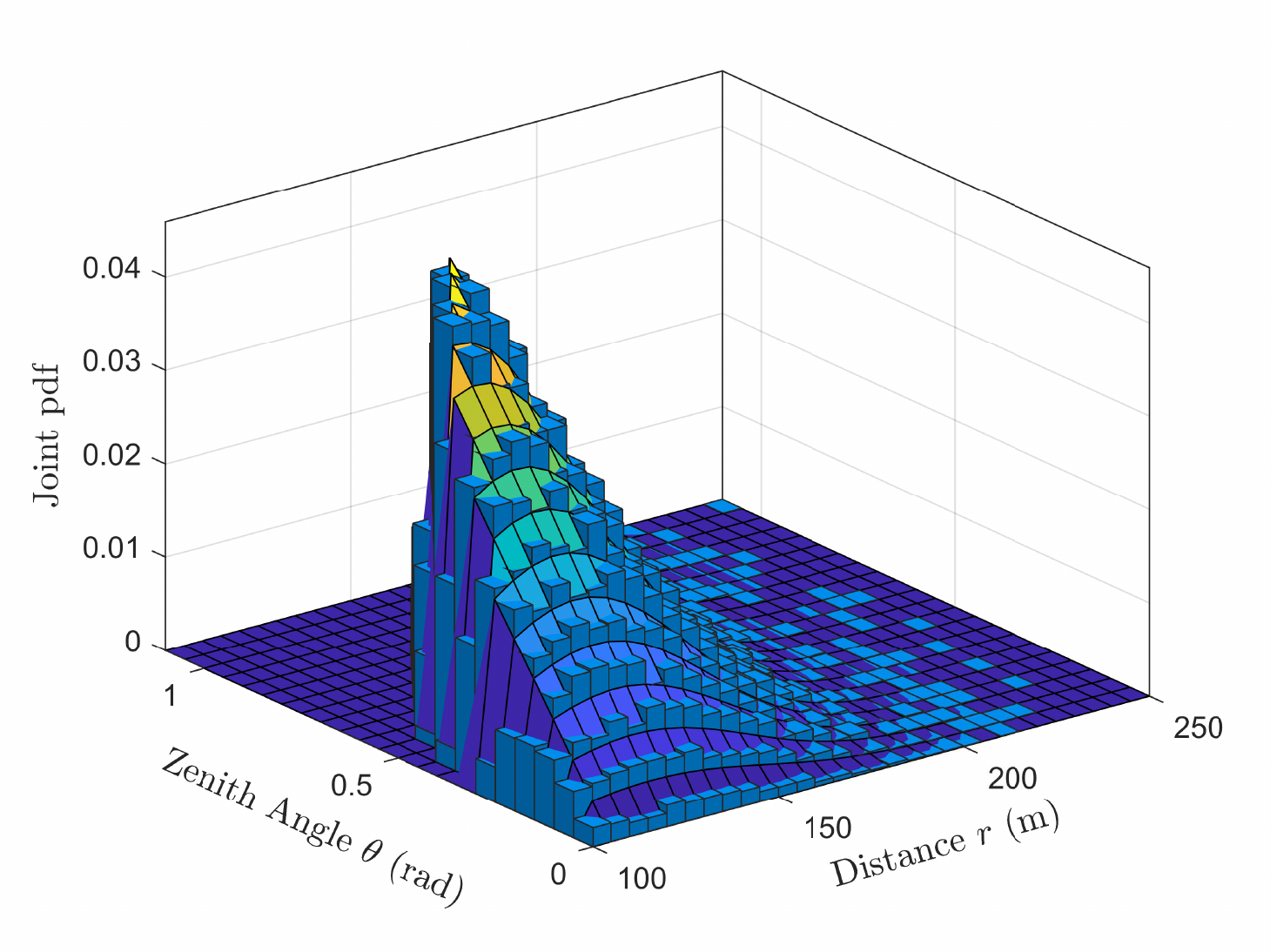}
\vspace{-0.7cm}
\caption{The joint pdf of $r_{\rm D_0, L}$ and $\theta_{\rm D_0, L}$ (suburban, $\lambda_{\rm D} = 10^{-6}$, $h_{\rm D, m} = 100$ m, and $h_{\rm D, M} = 300$ m).}
\vspace{-0.5cm}
\label{fig:JointPDF_r_theta}
\end{minipage}
\end{figure}
\vspace{-0.3cm}
\subsection{Intermediate Results}
We begin by focusing on the intermediate results given in Section \ref{sec:MathematicalConstructs}. In Fig. \ref{fig:CDF_T1T2T3_All}, we compare the analytical results for the cdf of $T_1$ (Lemma \ref{lem:01}), $T_2$ (Lemma \ref{lem:02}), and the joint cdf of $T_1$ and $T_3$ (Lemma \ref{lem:03}) with numerical simulations using representative parameters $a = 1$, $b = 4$, $I = 2$, $m = 2$, and $g = 1$. Since the coverage probability in the DF and AF relaying protocols is proportional to the ccdf of $T_2$ and the joint ccdf of $T_1$ and $T_3$, respectively, we can clearly observe the performance superiority of the DF over AF in this figure. The joint pdf of the serving distance ($r_{\rm D_0, L}$) and zenith angle ($\theta_{\rm D_0, L}$) when the serving UAV is in LoS is plotted in Fig. \ref{fig:JointPDF_r_theta} for a suburban environment with $\lambda_{\rm D} = 10^{-6}$, $h_{\rm D, m} = 100$ m, and $h_{\rm D, M} = 300$ m. Note that since the excessive path-loss is very high for an NLoS channel condition \cite{J_AlHourani_Optimal_2014}, the NLoS association probability will be low in such realistic environments and the closest LoS UAV to the origin almost always provides higher received power at the typical UE than the closest NLoS UAV. 

\vspace{-0.4cm}
\subsection{Impact of Relaying Protocols, UAV Height, and Density}
\begin{figure}[t!]
	\centering
	\begin{minipage}{0.32\columnwidth}
		\centering
		\includegraphics[width=\textwidth]{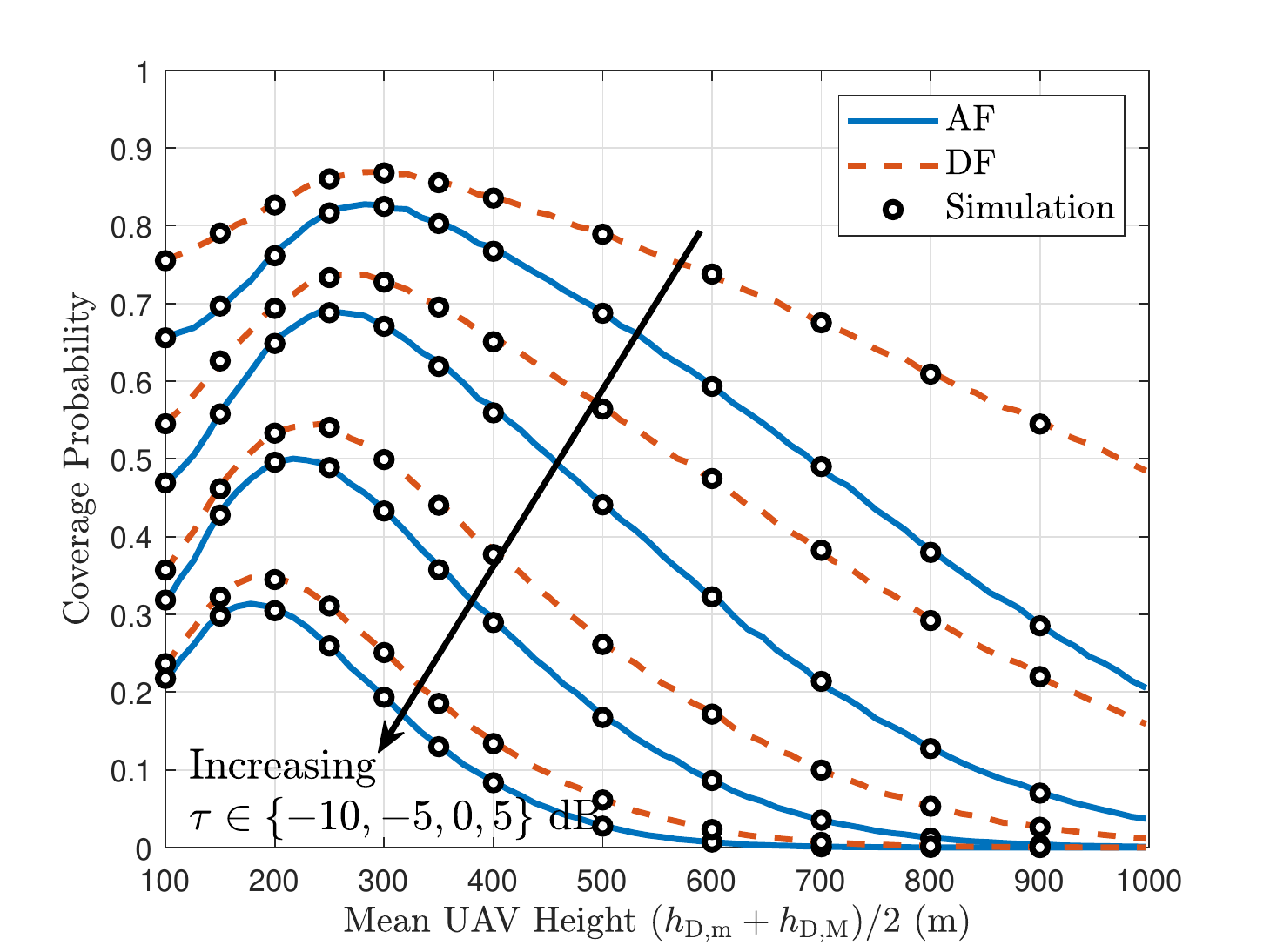}
		\vspace{-1.3cm}
		\caption{Coverage probability as a function of the mean UAV height for both AF and DF and different $\tau$'s \!(urban, $\lambda_{\rm D} \!=\! 10^{-8}$, $m \!=\! 1$, $h_{\rm D, M} - h_{\rm D, m} \!=\! 100$ m).}
		\vspace{-0.5cm}
		\label{fig:CompHeight1}
	\end{minipage}\hfill
	\begin{minipage}{0.32\columnwidth}
		\centering
		\includegraphics[width=\textwidth]{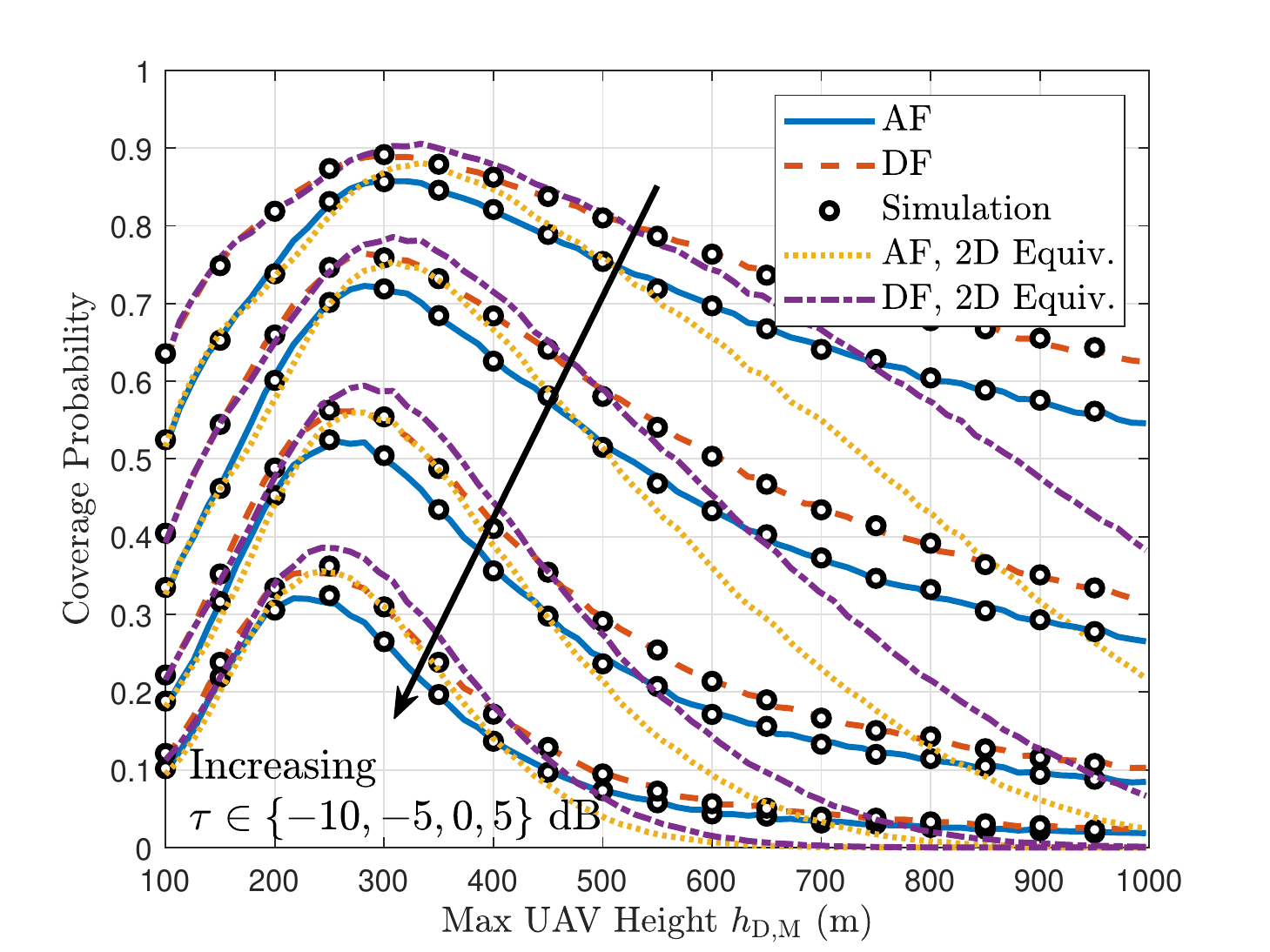}
		\vspace{-1.3cm}
		\caption{Coverage probability as a function of the maximum UAV height for both AF and DF and different $\tau$'s (urban, $\lambda_{\rm D} = 10^{-8}$, $m = 1$, $h_{\rm D, m} = 50$ m).}
		\vspace{-0.5cm}
		\label{fig:CompHeight2}
	\end{minipage}\hfill
	\begin{minipage}{0.32\columnwidth}
		\centering
		\includegraphics[width=\textwidth]{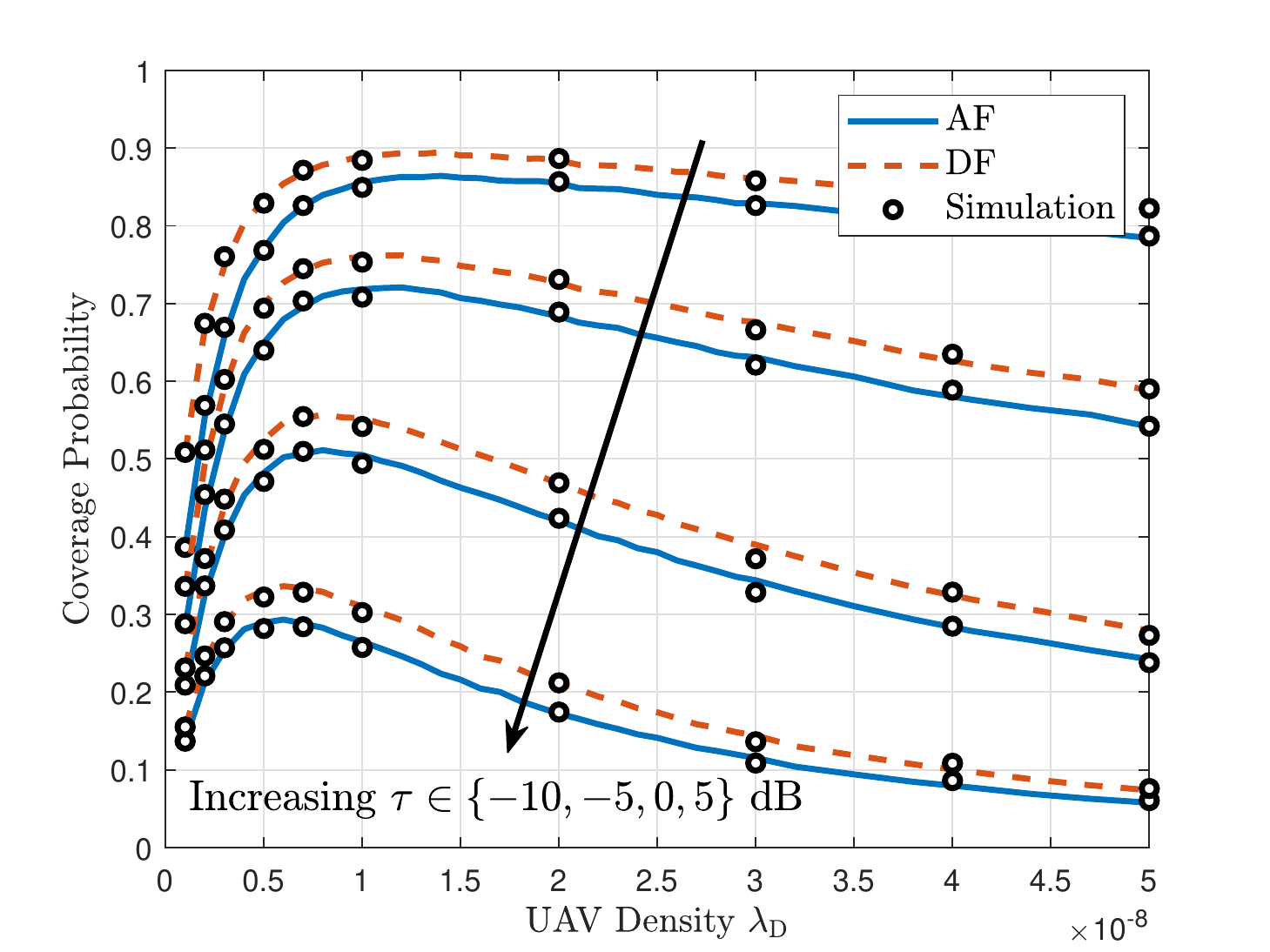}
		\vspace{-1.3cm}
		\caption{Coverage probability as a function of UAV density for both AF and DF and different $\tau$'s (urban, $m = 1$, $h_{\rm D, m} = 100$ m, $h_{\rm D, M} = 300$ m).}
		\vspace{-0.5cm}
		\label{fig:CompLam}
	\end{minipage}
\end{figure}

In Figs. \ref{fig:CompHeight1}-\ref{fig:CompLam}, we show the coverage probability as a function of UAV height and density for both AF and DF relaying protocols in an urban environment. In Fig. \ref{fig:CompHeight1}, we keep the difference between the maximum and minimum UAV heights constant ($h_{\rm D, M} - h_{\rm D, m} = 100$ m) and then increase the mean UAV height from $100$ m to $1000$ m. On the other hand, in Fig. \ref{fig:CompHeight2}, we keep the minimum UAV height constant ($h_{\rm D, m} = 50$ m) and increase the maximum UAV height from $100$ m to $1000$ m. In both of these figures, we set $\lambda_{\rm D} = 10^{-8}$ and $m = 1$. Let us define the equivalent 2D model of a 3D UAV network as a network with the following two properties: (i) all the UAVs are at the same height, which is set to be the mean value of the maximum and minimum UAV heights of the original 3D network, and (ii) the average number of points in the 2D model is the same as that of the 3D network (with the interpretation that all the points in the 3D setup are projected onto the 2D plane). Using this definition, we compare our results for the 3D network with its equivalent 2D model in Fig. \ref{fig:CompHeight2}. As seen from this figure, the two networks behave very similarly for small values of the height difference. However, as we increase the height difference, the coverage probability of the equivalent 2D model differs significantly from that of the 3D network. In Fig. \ref{fig:CompLam}, we assume $m = 1$, $h_{\rm D, m} = 100$ m, $h_{\rm D, M} = 300$ m, and obtain the coverage probability by increasing $\lambda_{\rm D}$ from $10^{-9}$ to $5 \times 10^{-8}$. The following observations can be made from these figures: (i) coverage probability in the DF relaying protocol is higher than that of the AF protocol, which has been theoretically shown in Section \ref{subsec5:RelayingMetric} and further pointed out in Fig. \ref{fig:CDF_T1T2T3_All}, (ii) coverage probability decreases as the SINR threshold $\tau$ increases, which is also clear from the definition, (iii) network performance can significantly benefit from limiting the maximum allowable UAV height, (iv) 3D UAV networks cannot always be accurately modeled using their equivalent 2D models, and (v) there exist mean UAV height $h_{\rm D}^*$ and UAV density $\lambda_{\rm D}^*$ for which the coverage probability is maximized for each SINR threshold $\tau$. Note that although increasing the average UAV height increases the LoS probability and makes the overall channel condition better, the increased UAV-UE distance significantly affects the path-loss and degrades the coverage probability. Furthermore, increasing the average number of UAVs per unit volume beyond $\lambda_{\rm D}^*$ increases the overall interference at the typical UE and the serving UAV, which in turn degrades the coverage probability.

\begin{figure}[t!]
\centering
\begin{minipage}{0.48\columnwidth}
\centering
\includegraphics[width=0.8\textwidth]{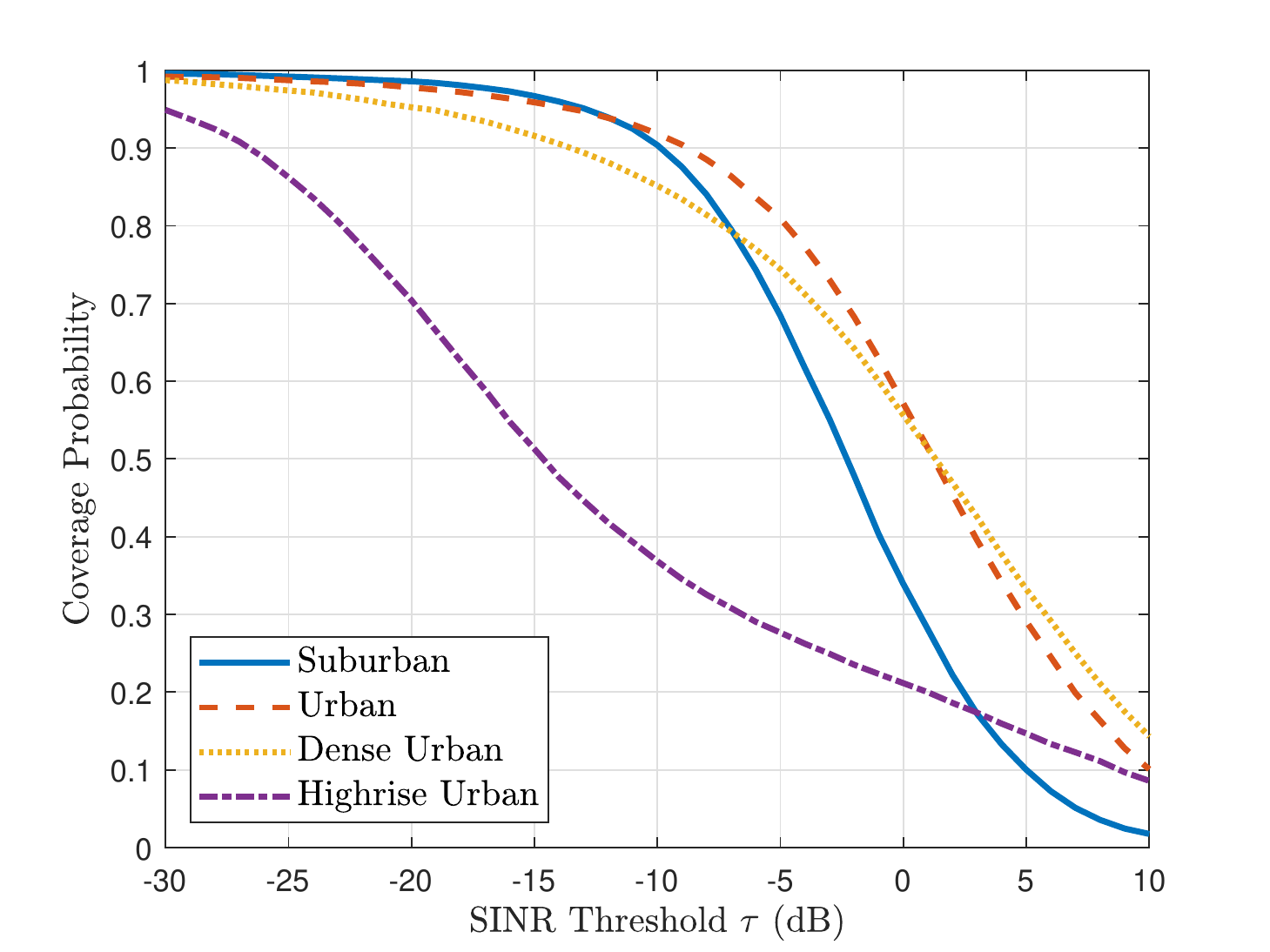}
\vspace{-0.7cm}
\caption{Coverage probability as a function of $\tau$ for the AF protocol at different environments ($\lambda_{\rm D} = 10^{-8}$, $m = 2$, $h_{\rm D, m} = 100$ m, $h_{\rm D, M} = 300$ m).}
\vspace{-0.5cm}
\label{fig:CompEnv_m2}
\end{minipage}\hfill
\begin{minipage}{0.48\columnwidth}
\centering
\includegraphics[width=0.8\textwidth]{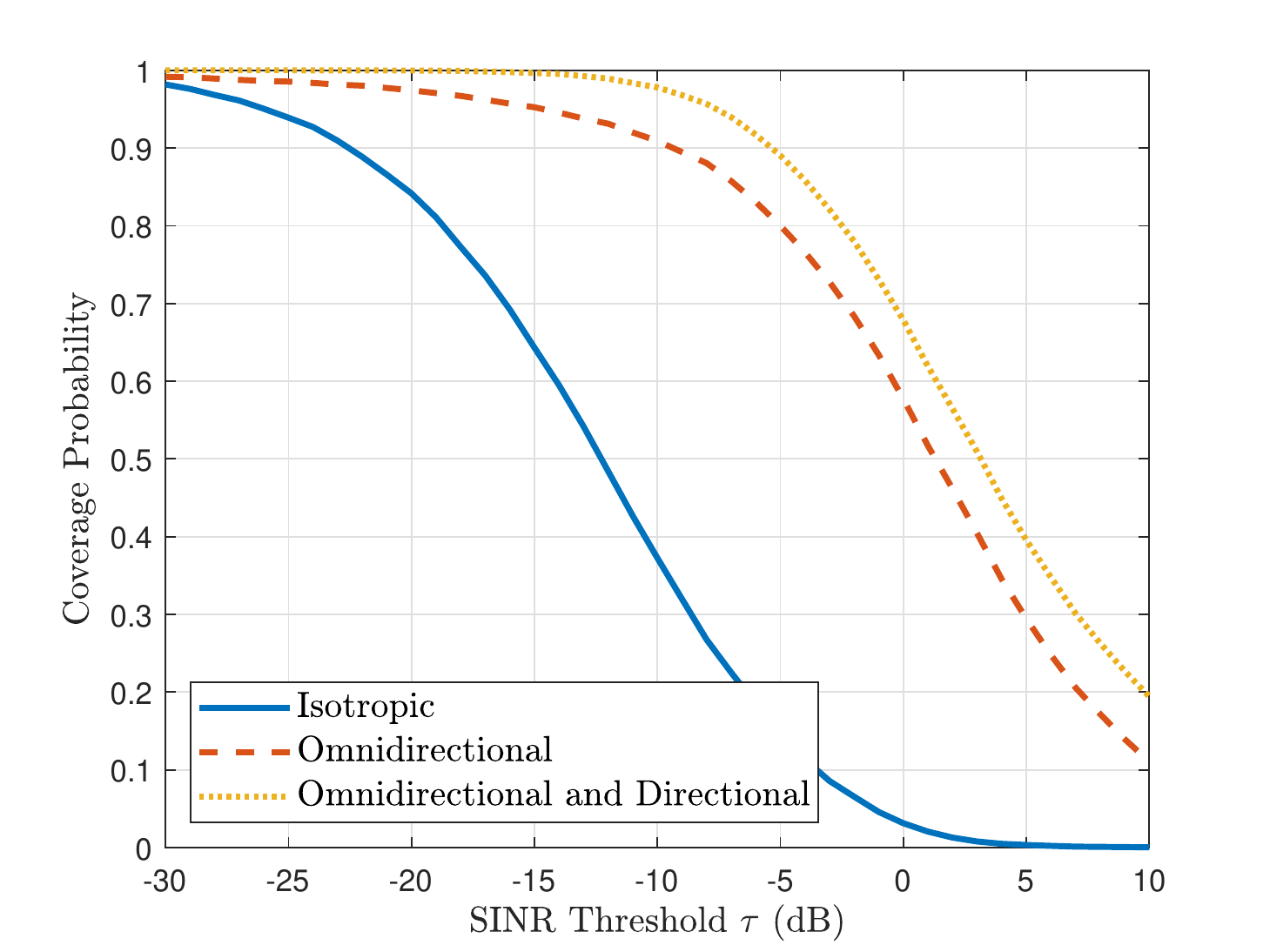}
\vspace{-0.7cm}
\caption{Coverage probability as a function of $\tau$ for the DF protocol using different BS antenna models (urban, $\lambda_{\rm D} = 10^{-8}$, $m = 2$, $h_{\rm D, m} = 100$ m, $h_{\rm D, M} = 300$ m).}
\vspace{-0.5cm}
\label{fig:CompAnt}
\end{minipage}
\end{figure}
\vspace{-0.3cm}

\subsection{Impact of Environments and Antenna Models}
We plot the AF coverage probability versus SINR threshold at different environments in Fig. \ref{fig:CompEnv_m2}, where the parameters are $\lambda_{\rm D} = 10^{-8}$, $m = 2$, $h_{\rm D, m} = 100$ m, and $h_{\rm D, M} = 300$ m. Due to high excessive path-loss and low LoS probability, we observe the worst coverage performance for most SINR thresholds in the highrise urban environment. It is also quite interesting to note that the suburban environment does not demonstrate the best performance. Although we have stronger received power at the typical UE in the suburban environment due to lower path-loss and higher LoS probability, the interference will also be stronger in this case. Therefore, numerical simulations are necessary to compare the coverage performance in different environments. In Fig. \ref{fig:CompAnt}, we show the DF coverage probability of the network in an urban environment for three antenna models: (i) \textit{isotropic}, where we assume all BSs and UAVs are equipped with isotropic antennas, (ii) \textit{omnidirectional}, where we assume each BS is equipped with one downtilted omnidirectional ULA, as described in Section \ref{par:OmniBS}, and (iii) \textit{omnidirectional and directional}, where each BS is equipped with one downtilted omnidirectional ULA and one uptilted directional antenna, as described in Section \ref{par:OmniDirBS}. The performance superiority of antenna models (ii) and (iii) over the canonical model in (i) is completely clear in this figure. We can also observe the benefit of having an uptilted antenna at the BS site that is solely used for backhaul purposes in this plot. With the antenna model given in (iii), the serving UAV is no longer served by the BS antenna sidelobes, which could be very weak. On the contrary, upon proper antenna orientation, the serving UAV will be served by a strong directional antenna at the serving BS, and thus, the BS-UAV backhaul link will be strong, which results in a better overall coverage performance.
\vspace{-0.2cm}
\section{Conclusion}
In this paper, we studied the performance of two-hop backhaul-aware 3D cellular networks, where BSs and UAVs coexist to serve the UEs on the ground. Specifically, each UE either connects directly to a fiber-backhauled terrestrial BS (access link), or connects first to a UAV which is then wirelessly backhauled to a terrestrial BS (joint access and backhaul). Inspired by the 3GPP studies, we used realistic antenna patterns for both BSs and UAVs. Due to the high probability of LoS in air-to-ground wireless communications, we adopted a probabilistic channel model for the UAV-UE links that incorporates both LoS and NLoS channel conditions. Following the max-power association policy, we characterized the network coverage performance for two well-known relaying protocols, i.e., AF and DF, by identifying and analyzing the building blocks of their SINR expressions. We also provided a comprehensive analysis of the joint distribution of distance and zenith angle of the closest and serving UAV to the typical UE in a 3D setting using tools from stochastic geometry. Moreover, since the UAV backhaul link could be much weaker than its access link due to the BS antenna sidelobes and nulls, we analyzed the addition of an uptilted directional antenna at the BS site for improving the UAV backhaul link. To the best of our understanding, this is the first work that offers a comprehensive analysis of 3D cellular networks where BSs provide wireless backhaul to the UAVs using a two-hop relaying scheme. While this work provides insightful results in a two-hop setting, extending its outcomes to a multi-hop scenario would be valuable, especially when coverage in faraway regions is required. Another possible extension of this work is the analysis of the two-hop transmission while incorporating spatial coupling in the placement of UAVs and UEs through the use of Poisson cluster processes \cite{J_Saha_3GPP_2018, J_Galkin_Stochastic_2019}. Furthermore, because we considered the joint transmission of BSs and UAVs to the UEs in this paper, studying their \textit{coordinated} joint transmission \cite{J_Rajanna_Downlink_2017} is a meaningful extension of this work. Since the serving BS and UAV do not interfere with each other's transmission in the coordinated scheme, we expect to get better results in terms of average rate and coverage probability.

\vspace{-0.4cm}
\appendix
\vspace{-0.4cm}
\subsection{Proof of Lemma \ref{lem:01}} \label{app:lem:01}
By definition, we have
\begin{align*}
	F_{T_1}(\tau) &= \nbbP\left[ \frac{aX}{bY + I} \leq \tau \right] = \nbbP\left[ aX \leq \tau b Y + \tau I \right] \overset{(a)}{=} \int_0^\infty f_Y(y) F_X\left(\frac{\tau b y + \tau I}{a}\right) \,{\rm d}y\\
	&\overset{(b)}{=} 1 - \sum_{i=0}^{m-1} \frac{m^m}{i! (m-1)!}\left(\frac{m\tau b}{a}\right)^i {\rm e}^{-\frac{m\tau I}{a}} \int_{0}^{\infty}y^{m-1}\left(y+\frac{I}{b}\right)^i {\rm e}^{-\left(m + \frac{m\tau b}{a}\right)y} \,{\rm d}y\\
	&\overset{(c)}{=} 1 - \sum_{i=0}^{m-1}\sum_{k=0}^{i} \frac{m^m (k+m-1)!}{i! (m-1)!} {i \choose k} \left(\frac{m\tau b}{a}\right)^i \left(\frac{I}{b}\right)^{i-k} \left(m + \frac{m\tau b}{a}\right)^{-m-k} {\rm e}^{-\frac{m\tau I}{a}},
\end{align*}
where in $(a)$ we used the independence of $X$ and $Y$, in $(b)$ we wrote the cdf of $X$ as the series expansion $F_X(x) = \frac{\gamma(m, mx)}{(m-1)!} = 1 - \sum_{i=0}^{m-1} \frac{(mx)^i}{i!}{\rm e}^{-mx}$ for integer $m$, and in $(c)$ we used the binomial expansion and simplified the resulting integral using the definition of the gamma function. By further mathematical manipulations, we obtain the final result as given in \eqref{CDF1}.
\hfill 
\IEEEQED
\vspace{-0.6cm}
\subsection{Proof of Lemma \ref{lem:02}} \label{app:lem:02}
Similar to the proof of Lemma \ref{lem:01}, we can write
\begin{align*}
	F_{T_2}(\tau) &= \nbbP\left[ \frac{\max\{aX, bY\}}{\min\{aX, bY\} \!+\! I} \leq \tau \right] \!= \nbbP\left[ \frac{aX}{bY \!+ I} \leq \tau, aX \geq bY \right] \!+ \nbbP\left[ \frac{bY}{aX \!+ I} \leq \tau, aX < bY \right]\\
	&= \begin{cases}
		\int_0^{\frac{\tau I}{(1-\tau)b}} \int_{\frac{by}{a}}^{\frac{\tau by + \tau I}{a}} f_{X,Y}(x, y)\,{\rm d}x\,{\rm d}y  +  \int_0^{\frac{\tau I}{(1-\tau)a}} \int_{\frac{ax}{b}}^{\frac{\tau ax + \tau I}{b}} f_{X,Y}(x, y)\,{\rm d}y\,{\rm d}x & \tau < 1\\
		\int_0^{\infty} \int_{\frac{by}{a}}^{\frac{\tau by + \tau I}{a}} f_{X,Y}(x, y)\,{\rm d}x\,{\rm d}y  +  \int_0^{\infty} \int_{\frac{ax}{b}}^{\frac{\tau ax + \tau I}{b}} f_{X,Y}(x, y)\,{\rm d}y\,{\rm d}x & \tau \geq 1
	\end{cases}.
\end{align*}
We now focus our attention to region $\tau < 1$ and denote its first double integral as $L_1$. We have
\begin{align*}
	L_1 &\overset{(a)}{=} \int_0^{\frac{\tau I}{(1-\tau)b}} \frac{m^m}{(m-1)!}y^{m-1}{\rm e}^{-my} \sum_{i=0}^{m-1}\frac{1}{i!} \left[ \left(\frac{mb}{a}y\right)^{i} {\rm e}^{-\frac{mb}{a}y} - \left(\frac{m\tau b}{a}y + \frac{m\tau}{a}I\right)^{i} {\rm e}^{-\frac{m\tau b}{a}y - \frac{m\tau}{a}I} \right] {\rm d}y\\
	&\overset{(b)}{=} \sum_{i=0}^{m-1} \frac{m^m}{i! (m-1)!} \left(\frac{mb}{a}\right)^{i} \int_{0}^{\frac{\tau I}{(1-\tau)b}} y^{i+m-1}{\rm e}^{-\left(m + \frac{mb}{a}\right)y} \,{\rm d}y\\
	&\quad- \sum_{i=0}^{m-1} \frac{m^m}{i! (m-1)!}\left(\frac{m\tau b}{a}\right)^i {\rm e}^{-\frac{m\tau}{a}I} \int_{0}^{\frac{\tau I}{(1-\tau)b}}y^{m-1}\left(y+\frac{I}{b}\right)^i {\rm e}^{-\left(m + \frac{m\tau b}{a}\right)y} \,{\rm d}y\\
	&\overset{(c)}{=} \sum_{i=0}^{m-1} \frac{\gamma\left(m+i, \left(\frac{1}{a} + \frac{1}{b}\right)\frac{m\tau}{(1-\tau)}I\right)}{i!(m-1)!} \frac{a^m b^i}{(a + b)^{m + i}}\\
	&\quad- \sum_{i=0}^{m-1}\sum_{k=0}^{i} \frac{\gamma\left(m+k, \left(\frac{\tau}{a} + \frac{1}{b}\right)\frac{m\tau}{(1-\tau)}I\right)} {k!(m-1)!(i-k)!} \frac{a^m (b\tau)^k}{(a + b\tau)^{m + k}} \left(\frac{m\tau}{a}I\right)^{i-k} {\rm e}^{-\frac{m\tau}{a}I}
\end{align*}
where in $(a)$ we used the independence between $X$ and $Y$ and the series expansion of the cdf of gamma random variables (as given in the proof of Lemma \ref{lem:01}), in $(b)$ we switched the order of summation and integration, and in $(c)$ we derived the integrals and simplified the resulting expressions using the binomial expansion and the definition of the lower incomplete gamma function. Note that the second double integral is nothing but $L_1$ with $a$ and $b$ being switched with each other. As for region $\tau \geq 1$, the proof follows the same steps as above, with the only difference that the upper limits of the outer integrals are $\infty$ and we end up with gamma functions instead of incomplete gamma functions in the last step. Noting that $\gamma(s, \infty) = \Gamma(s)$, we obtain the final result as given in \eqref{CDF2}.
\hfill 
\IEEEQED
\vspace{-0.5cm}
\subsection{Proof of Lemma \ref{lem:03}} \label{app:lem:03}
We start by writing the joint cdf of interest as
\begin{align}\label{eq1_lem03}
	F_{T_1, T_3}(\tau, \tau) &= \nbbP\left[ \frac{aX}{bY + I} \leq \tau, \frac{bY}{aX + I + g(aX + bY + I)} \leq \tau \right]\nonumber\\
	&= \nbbP\left[ bY \geq \frac{a}{\tau}X - I, (1-\tau g)bY \leq \tau(1+g)aX + \tau(1+g)I \right].
\end{align}
When $\tau \geq \frac{1}{g}$, the second condition in \eqref{eq1_lem03} always holds since all the constants and random variables on its right-hand side are non-negative. Hence, we have $F_{T_1, T_3}(\tau, \tau) = \nbbP\left[ \frac{aX}{bY + I} \leq \tau \right] = F_{T_1}(\tau)$ for $\tau \geq \frac{1}{g}$. When $\tau < \frac{1}{g}$, we first find the intersection point of lines $Y = \frac{a}{b\tau}X - \frac{1}{b}I$ and $Y = \frac{a\tau(1+g)}{b(1-\tau g)}X + \frac{\tau(1+g)}{b(1-\tau g)}I$ as $x_0 = \frac{\tau I}{a(1-\tau(1+g))}, y_0 = \frac{\tau(1+g) I}{b(1-\tau(1+g))}$.
Note also that $by_0 = (1+g)ax_0$. We can now derive $F_{T_1, T_3}(\tau, \tau)$ by integrating over the region defined in \eqref{eq1_lem03} as follows:
\begin{align*}
	F_{T_1, T_3}(\tau, \tau) &= \begin{cases}
		\int_0^{y_0} \!\!\int_{\frac{by}{a(1+g)}}^{\frac{\tau by + \tau I}{a}} \!f_{X,Y}(x, y)\,{\rm d}x\,{\rm d}y  +  \int_0^{x_0}\!\! \int_{\frac{(1+g)ax}{b}}^{\frac{\tau (1+g)}{b(1-\tau g)}(ax + I)} \!f_{X,Y}(x, y)\,{\rm d}y\,{\rm d}x & \tau < \frac{1}{1+g}\\
		\int_0^{\infty}\!\! \int_{\frac{by}{a(1+g)}}^{\frac{\tau by + \tau I}{a}} \!f_{X,Y}(x, y)\,{\rm d}x\,{\rm d}y  +  \int_0^{\infty}\!\! \int_{\frac{(1+g)ax}{b}}^{\frac{\tau (1+g)}{b(1-\tau g)}(ax + I)} \!f_{X,Y}(x, y)\,{\rm d}y\,{\rm d}x & \frac{1}{1+g} \leq \tau < \frac{1}{g}
	\end{cases}.
\end{align*}
This equation can be simplified similarly as in the proof of Lemma \ref{lem:02} to obtain \eqref{CDF3}. These steps are skipped here for brevity. This completes the proof.
\hfill 
\IEEEQED
\vspace{-0.5cm}
\subsection{Proof of Lemma \ref{lem:3}} \label{app:lem:3}
From \eqref{Assoc_UAV} and for a \textit{specific} channel condition, we observe that among all the UAVs, the closest one to the typical UE provides the highest averaged received power. Hence, the serving UAV will be either the closest LoS UAV or the closest NLoS UAV to the typical UE. We can now write the probability that an NLoS UAV serves the typical UE as
\begin{align*}
	A_{\rm N} \!=\! \nbbP\left[ P_{\rm D_0, N}^{\rm Rx}\! \geq\! P_{\rm D_0, L}^{\rm Rx} \right] \!=\! \nbbP\left[ \tilde{r}_{{\rm D_0, N}}^{-\alpha_{\rm N}} \eta_{\rm N}^{-1} \!\geq\! \tilde{r}_{{\rm D_0, L}}^{-\alpha_{\rm L}} \eta_{\rm L}^{-1} \right] \!\overset{(a)}{=}\! \int_{h_{\rm D, m}}^{\infty} \!\!\nbbP\left[ \tilde{r}_{{\rm D_0, L}} \geq\! \left(\frac{\eta_{\rm N}}{\eta_{\rm L}}\right)^{\frac{1}{\alpha_{\rm L}}} \!r^{\frac{\alpha_{\rm N}}{\alpha_{\rm L}}} \right] \!f_{\tilde{r}_{{\rm D_0, N}}}(r) \,{\rm d}r,
\end{align*}
where in $(a)$ we used the law of total probability by conditioning on $\tilde{r}_{{\rm D_0, N}}$. The final result in \eqref{Association_L_D} is derived by applying the complementary cdf (ccdf) and pdf of $\tilde{r}_{{\rm D_0, L}}$ and $\tilde{r}_{{\rm D_0, N}}$, respectively, from \eqref{Dist_rTilde_D} to the last integral above. Since the typical UE associates with either an LoS UAV or an NLoS UAV, we have $A_{\rm L} = 1 - A_{\rm N}$.
\hfill 
\IEEEQED
\vspace{-0.5cm}
\subsection{Proof of Lemma \ref{lem:4}} \label{app:lem:4}
Considering the LoS case, we write the joint cdf of $r_{\rm D_0, L}$ and $\theta_{\rm D_0, L}$ as $F_{r_{\rm D_0, L}, \theta_{\rm D_0, L}}(r, \theta)$
\begin{align*}
	&= \nbbP\left[ \scalebox{0.95}{$\tilde{r}_{\rm D_0, L} \leq r, \tilde{\theta}_{\rm D_0, L} \leq \theta \,|\, {\rm D_0 ~ is ~ an ~ LoS ~ UAV}$} \right] \!\overset{(a)}{=}\! \frac{1}{A_{\rm L}} \nbbP\left[ \scalebox{0.95}{$\tilde{r}_{\rm D_0, L} \leq r, \tilde{\theta}_{\rm D_0, L} \leq \theta, \tilde{r}_{{\rm D_0, L}}^{-\alpha_{\rm L}} \eta_{\rm L}^{-1} \geq \tilde{r}_{{\rm D_0, N}}^{-\alpha_{\rm N}} \eta_{\rm N}^{-1}$} \right]\\
	&= \frac{1}{A_{\rm L}} \int_{0}^{\theta} \int_{h_{\rm D, m}}^r \nbbP\left[ \tilde{r}_{{\rm D_0, N}} \geq \left(\frac{\eta_{\rm L}}{\eta_{\rm N}}\right)^{\frac{1}{\alpha_{\rm N}}} r_1^{\frac{\alpha_{\rm L}}{\alpha_{\rm N}}} \right] f_{\tilde{r}_{{\rm D_0, L}}, \tilde{\theta}_{\rm D_0, L}}(r_1, \theta_1) \,{\rm d}r_1 \,{\rm d}\theta_1\\
	&\overset{(b)}{=} \frac{1}{A_{\rm L}} \int_{0}^{\theta} \int_{h_{\rm D, m}}^{\min\{r, r_0\}} f_{\tilde{r}_{{\rm D_0, L}}, \tilde{\theta}_{\rm D_0, L}}(r_1, \theta_1) \,{\rm d}r_1 \,{\rm d}\theta_1 \\
	&\quad + \frac{1}{A_{\rm L}} \int_{0}^{\theta} \int_{\min\{r, r_0\}}^r \exp\left[-\pi \lambda_{\rm D} \beta_{\rm N}\left( \left(\frac{\eta_{\rm L}}{\eta_{\rm N}}\right)^{\frac{1}{\alpha_{\rm N}}} r_1^{\frac{\alpha_{\rm L}}{\alpha_{\rm N}}} \right)\right] f_{\tilde{r}_{{\rm D_0, L}}, \tilde{\theta}_{\rm D_0, L}}(r_1, \theta_1) \,{\rm d}r_1 \,{\rm d}\theta_1,
\end{align*}
where $r_0 = (\eta_{\rm N}/\eta_{\rm L})^{1/\alpha_{\rm L}} h_{\rm D, m}^{\alpha_{\rm N}/\alpha_{\rm L}}$, in $(a)$ we used the fact that $A_{\rm L} = \nbbP\left[ {\rm D_0 ~ is ~ an ~ LoS ~ UAV} \right]$, and in $(b)$ we used $\alpha_{\rm L} < \alpha_{\rm N}$, $\eta_{\rm L} < \eta_{\rm N}$, and the ccdf of $\tilde{r}_{{\rm D_0, N}}$ from \eqref{Dist_rTilde_D}. Taking the derivative of $F_{r_{\rm D_0, L}, \theta_{\rm D_0, L}}(r, \theta)$ with respect to both $r$ and $\theta$, we obtain the final result as given in \eqref{Serving_L_D}.
\hfill 
\IEEEQED
\vspace{-0.5cm}
\subsection{Proof of Lemma \ref{lem:5}} \label{app:lem:5}
By definition, we have
\begin{align*}
	\ncalL_{I_{\rm BU}}(s|{\rm B_0}) &= \nbbE\left[ {\rm e}^{-s \sum_{{\rm B}_\nbx \in \Phi_{\rm B}'}  P_{\rm B} \eta_{\rm N}^{-1} G_{{\rm B}_\nbx} r_{{\rm B}_\nbx}^{-\alpha_{\rm N}} f_{{\rm B}_\nbx}} \middle|{\rm B_0}\right] \overset{(a)}{=} \nbbE\left[ \prod_{{\rm B}_\nbx \in \Phi_{\rm B}'} \left( 1 + \frac{sP_{\rm B}}{m\eta_{\rm N}}\frac{G_{{\rm B}_\nbx}}{r_{{\rm B}_\nbx}^{\alpha_{\rm N}}} \right)^{\!\!-m} \middle|{\rm B_0}\right]\\
	&\overset{(b)}{=} \exp\left[ -2\pi\lambda_{\rm B}\int_{u_{\rm B_0}}^{\infty} \left[ 1 - \left( 1 + \frac{sP_{\rm B}}{m\eta_{\rm N}}\frac{G_{{\rm B}_\nbx}}{r_{{\rm B}_\nbx}^{\alpha_{\rm N}}} \right)^{\!\!-m} \right] u_{{\rm B}_\nbx} {\rm d}u_{{\rm B}_\nbx} \right],
\end{align*}
where in $(a)$ we took the moment generating function (mgf) of the gamma-distributed $f_{{\rm B}_\nbx}$, and in $(b)$ we used the probability generating functional (pgfl) of the PPP $\Phi_{\rm B}'$. Noting that $r_{{\rm B}_\nbx} = \sqrt{u_{{\rm B}_\nbx}^2 + h_{\rm B}^2}$ and $G_{{\rm B}_\nbx} = G_{\rm B}^{\rm OmniD}\left(\pi - \theta_{{\rm B}_\nbx}, \theta_{\rm B}\right)$, where $\theta_{{\rm B}_\nbx} = \tan^{-1}(\frac{u_{{\rm B}_\nbx}}{h_{\rm B}})$ is the zenith angle of the BS located at ${\rm B}_\nbx$, the final result in \eqref{Lap_I_BU} is obtained.
\hfill 
\IEEEQED
\vspace{-0.5cm}
\subsection{Proof of Lemma \ref{lem:6}} \label{app:lem:6}
Similar to the proof of Lemma \ref{lem:5}, we have
\begin{align*}
	&\ncalL_{I_{{\rm DU}, q_1|q_2}}\!(s|{\rm D_0}) \!= \nbbE\left[ {\rm e}^{-s \sum_{{\rm D}_\nbx \in \Phi_{{\rm D}, q_1}'} \!\!\! P_{\rm D} \eta_{q_1}^{-1} G_{{\rm D}_\nbx} r_{{\rm D}_\nbx}^{-\alpha_{q_1}} f_{{\rm D}_\nbx}} \middle|{\rm D_0}\right] \!\!= \nbbE\left[ \prod_{{\rm D}_\nbx \in \Phi_{{\rm D}, q_1}'}\!\!\!\! \left( \!1 \!+\! \frac{sP_{\rm D}}{m\eta_{q_1}}\frac{G_{{\rm D}_\nbx}}{r_{{\rm D}_\nbx}^{\alpha_{q_1}}} \!\right)^{\!\!-m} \middle|{\rm D_0}\right]\\
	&= \!\exp\!\Bigg[\! -\int_0^{2\pi}\!\!\int_{\theta_{\rm D, M}}^{\theta_{\rm D, m}}\! \int_{r_{q_1|q_2}}^{\frac{h_{\rm D, M}}{\cos(\theta_{{\rm D}_\nbx})}} \!\bigg[ 1 \!- \!\Big( 1 \!+\! \frac{sP_{\rm D}}{m\eta_{q_1}}\frac{G_{{\rm D}_\nbx}}{r_{{\rm D}_\nbx}^{\alpha_{q_1}}} \Big)^{\!\!-m} \bigg] \lambda_{\rm D}p_{q_1}(\theta_{{\rm D}_\nbx}) r_{{\rm D}_\nbx}^2\!\sin(\theta_{{\rm D}_\nbx}) {\rm d}r_{{\rm D}_\nbx} {\rm d}\theta_{{\rm D}_\nbx} {\rm d}\phi_{{\rm D}_\nbx} \\
	&\hspace{1.36cm}-\int_0^{2\pi}\!\!\int_{\theta_{\rm D, m}}^{\frac{\pi}{2}}\! \int_{\frac{h_{\rm D, m}}{\cos(\theta_{{\rm D}_\nbx})}}^{\frac{h_{\rm D, M}}{\cos(\theta_{{\rm D}_\nbx})}} \!\bigg[ 1 \!- \!\Big( 1 \!+\! \frac{sP_{\rm D}}{m\eta_{q_1}}\frac{G_{{\rm D}_\nbx}}{r_{{\rm D}_\nbx}^{\alpha_{q_1}}} \Big)^{\!\!-m} \bigg] \lambda_{\rm D}p_{q_1}(\theta_{{\rm D}_\nbx}) r_{{\rm D}_\nbx}^2\!\sin(\theta_{{\rm D}_\nbx}) {\rm d}r_{{\rm D}_\nbx} {\rm d}\theta_{{\rm D}_\nbx} {\rm d}\phi_{{\rm D}_\nbx} \Bigg],
\end{align*}
where $\theta_{\rm D, m} = \cos^{-1}(\frac{h_{\rm D, m}}{r_{q_1|q_2}})$, $\theta_{\rm D, M} = \cos^{-1}\left(\min\left\{\frac{h_{\rm D, M}}{r_{q_1|q_2}}, 1\right\}\right)$, and we used the pgfl of the PPP $\Phi_{{\rm D}, q_1}'$ in the last equation. Note that the triple integration is carried out over the region enclosed between the planes $z = h_{\rm D, m}$ and $z = h_{\rm D, M}$ minus the exclusion zone $\ncalX_{\rm D}$ of the interfering UAVs, as explained earlier with details in Table \ref{table:ExcZone}. Observing that $G_{{\rm D}_\nbx} = G_{\rm D}^{\rm AC}\left(\pi - \theta_{{\rm D}_\nbx}\right)$, where $\theta_{{\rm D}_\nbx}$ is the zenith angle of the UAV located at ${\rm D}_\nbx$, we obtain the final result as given in \eqref{Lap_I_DU}.
\hfill 
\IEEEQED
\vspace{-0.5cm}
\subsection{Proof of Theorem \ref{theorem1}} \label{app:theorem1}
We start by writing the definition of the coverage probability as
\begin{align*}
	P_{\rm Cov}^{\rm AF} &= \nbbP\left[ {\rm SINR^{AF}} \geq \tau \right] = \nbbP\left[ {\rm SINR^{AF}} \geq \tau \middle| E_{\rm L} \right] \nbbP\left[ E_{\rm L} \right] + \nbbP\left[ {\rm SINR^{AF}} \geq \tau \middle| E_{\rm N} \right] \nbbP\left[ E_{\rm N} \right]\\
	&=\nbbP\left[ {\rm SINR_{\rm L}^{AF}} \geq \tau \right] A_{\rm L} + \nbbP\left[ {\rm SINR_{N}^{AF}} \geq \tau \right] A_{\rm N},
\end{align*}
where $E_{\rm L}$ and $E_{\rm N}$ represent the events that the typical UE is associated with an LoS and NLoS UAV, respectively, with probabilities of $A_{\rm L}$ and $A_{\rm N}$. We now write the ccdf of ${\rm SINR}_q^{\rm AF}$ as
\begin{align}\label{eq3:thm1}
	P_{{\rm Cov}, q}^{\rm AF} &\triangleq \nbbP\left[ {\rm SINR}_q^{\rm AF} \geq \tau \right] = 1 - \nbbP\left[ {\rm SINR}_{{\rm BU}, q} < \tau,  \frac{{\rm SINR_{BD}} {{\rm SINR}_{{\rm DU}, q}}}{{\rm SINR_{BD}} + {{\rm SINR}_{{\rm DU}, q}} + 1} < \tau \right].
\end{align}
Recall that conditioned on knowing the locations of the serving BS and UAV, the SINR values in \eqref{eq3:thm1} can be represented in simpler forms as given in \eqref{SINR2}. Using this representation and by further conditioning on $I$ and $Z$, we have
\begin{align*}
	P_{{\rm Cov}, q}^{\rm AF} = 1 - \nbbE\left[ \nbbP\left[ \frac{aX}{bY \!+\! I} \!<\! \tau,  \frac{bY}{aX \!+\! I \!+\! g(aX \!+\! bY \!+\! I)} \!<\! \tau \middle| {\rm B}_0, {\rm D}_0, I, Z \right] \right] \!= \nbbE\left[ \bar{F}_{T_1, T_3}\!\left( \tau, \tau \right) \right],
\end{align*}
where $g = \frac{N_0}{cZ}$ and we used Lemma \ref{lem:03} in the last equation. Note that $\bar{F}_{T_1, T_3}\!\left( \tau, \tau \right) = 1 - F_{T_1, T_3}\!\left( \tau, \tau \right)$ is the joint ccdf of $T_1$ and $T_3$ conditioned on knowing $a$, $b$, $g$, and $I$, which entails the incomplete gamma function, and thus, the series expansion $\gamma(s, x) = (s-1)!\left[1- \sum_{j=0}^{s-1} \frac{x^j}{j!}{\rm e}^{-x}\right]$ can be used to obtain $P_{{\rm Cov}, q}^{\rm AF}$. Taking the expectation of $\bar{F}_{T_1, T_3}\!\left( \tau, \tau \right)$ over $I = I_{\rm U} + N_0$, we end up with the derivatives of the Laplace transform of $I$, where $\ncalL_{I}(s|{\rm B_0}, {\rm D_0}) = \nbbE\left[ {\rm e}^{-s(I_{\rm U} + N_0)} \middle| {\rm B_0}, {\rm D_0} \right] = {\rm e}^{-sN_0} \ncalL_{I_{{\rm U}|q}}(s|{\rm B_0}, {\rm D_0})$. Following Remark \ref{remark5}, Lemma \ref{lem:03}, and Lemma \ref{lem:01}, we denote $\nbbE_I\left[ \bar{F}_{T_1, T_3}\!\left( \tau, \tau \right) \right]$ by $W_i$ for region $\hat{\ncalR}_i$, $i\in\{1, 2, 3\}$, as given in the theorem statement. Since $\hat{\ncalR}_i$ partitions the whole space, we obtain the final result as given in \eqref{eq2:thm1} by taking the expectation of each $W_i$ over the joint distribution of $Z$, $r_{{\rm B}_0}$, $\theta_{{\rm D}_0, q}$, $r_{{\rm D}_0, q}$, and $\phi_{\rm B_0 D_0}$.
\hfill 
\IEEEQED
\vspace{-0.5cm}
\subsection{Proof of Theorem \ref{theorem2}} \label{app:theorem2}
Similar to the proof of Theorem \ref{theorem1}, we only need to derive $P_{{\rm Cov}, q}^{\rm DF}$. We have
\begin{align*}
	P_{{\rm Cov}, q}^{\rm DF} &\triangleq \nbbP\left[ {\rm SINR}_q^{\rm DF} \geq \tau \right] = 1 - \nbbP\left[ {\rm SINR}_{{\rm BU}, q} < \tau,  \min\left\{ {\rm SINR_{BD}}, {{\rm SINR}_{{\rm DU}, q}} \right\} < \tau \right]\\
	&= 1 - \nbbP\left[ {\rm SINR}_{{\rm BU}, q} < \tau \right] + \nbbP\left[ {\rm SINR}_{{\rm BU}, q} < \tau,  {\rm SINR_{BD}} \geq \tau, {{\rm SINR}_{{\rm DU}, q}} \geq \tau \right]\\
	&\overset{(a)}{=} 1 - \nbbE\big[ \nbbP\left[ {\rm SINR}_{{\rm BU}, q} < \tau \middle| {\rm B}_0, {\rm D}_0, I \right] \big]\\
	&\qquad +  \nbbE\big[\nbbP\left[ {\rm SINR}_{{\rm BU}, q} < \tau, {{\rm SINR}_{{\rm DU}, q}} \geq \tau \middle| {\rm B}_0, {\rm D}_0, I \right]\nbbP\left[ {\rm SINR_{BD}} \geq \tau \middle| {\rm B}_0, {\rm D}_0 \right] \big]\\
	&\overset{(b)}{=} 1 - \nbbE\left[ \nbbP\left[ \frac{aX}{bY + I} < \tau \middle| {\rm B}_0, {\rm D}_0, I \right] \nbbP\left[ \frac{cZ}{N_0} < \tau \middle| {\rm B}_0, {\rm D}_0 \right] \right]\\
	&\qquad - \nbbE\left[ \nbbP\left[  \frac{\max\{aX, bY\}}{\min\{aX, bY\} + I} < \tau \middle| {\rm B}_0, {\rm D}_0, I \right] \nbbP\left[ \frac{cZ}{N_0} \geq \tau \middle| {\rm B}_0, {\rm D}_0 \right] \right],
\end{align*}
where in $(a)$ we conditioned the probabilities on knowing ${\rm B}_0$, ${\rm D}_0$, and $I = I_{\rm U} + N_0$, and in $(b)$ we wrote the SINR values in their simpler forms as in \eqref{SINR2} and also used the relation $\nbbP\left[ E \cap F \right] + \nbbP\left[ E \cap \bar{F} \right] = \nbbP\left[ E \right]$, where $E = {\rm SINR}_{{\rm BU}, q}$ and $F = {{\rm SINR}_{{\rm DU}, q}}$, to further simplify the result. Now, using Lemmas \ref{lem:01} and \ref{lem:02} and deconditioning similar to the proof of Theorem \ref{theorem1}, we obtain the final result as given in \eqref{eq2:thm2}.
\hfill 
\IEEEQED
\newcommand{\BIBdecl}{\setlength{\itemsep}{-0.225 em}}

\bibliographystyle{IEEEtran}
\bibliography{J4_3D_Backhaul_Drone}

\begin{thebibliography}{10}
\providecommand{\url}[1]{#1}
\csname url@samestyle\endcsname
\providecommand{\newblock}{\relax}
\providecommand{\bibinfo}[2]{#2}
\providecommand{\BIBentrySTDinterwordspacing}{\spaceskip=0pt\relax}
\providecommand{\BIBentryALTinterwordstretchfactor}{4}
\providecommand{\BIBentryALTinterwordspacing}{\spaceskip=\fontdimen2\font plus
\BIBentryALTinterwordstretchfactor\fontdimen3\font minus
  \fontdimen4\font\relax}
\providecommand{\BIBforeignlanguage}[2]{{%
\expandafter\ifx\csname l@#1\endcsname\relax
\typeout{** WARNING: IEEEtran.bst: No hyphenation pattern has been}%
\typeout{** loaded for the language `#1'. Using the pattern for}%
\typeout{** the default language instead.}%
\else
\language=\csname l@#1\endcsname
\fi
#2}}
\providecommand{\BIBdecl}{\relax}
\BIBdecl

\bibitem{C_Morteza_Fundamentals_2021}
M.~Banagar and H.~S. Dhillon, ``Fundamentals of {3D} two-hop cellular networks
  analysis with wireless backhauled {UAVs},'' in \emph{Proc. IEEE Global
  Commun. Conf. (Globecom)}, Dec. 2021.

\bibitem{J_Zeng_Wireless_2016}
Y.~Zeng, R.~Zhang, and T.~J. Lim, ``Wireless communications with unmanned
  aerial vehicles: Opportunities and challenges,'' \emph{IEEE Commun. Mag.},
  vol.~54, no.~5, pp. 36--42, May 2016.

\bibitem{J_Mozaffari_Tutorial_2018}
M.~{Mozaffari} \emph{et~al.}, ``A tutorial on {UAV}s for wireless networks:
  Applications, challenges, and open problems,'' \emph{IEEE Commun. Surveys
  Tuts.}, vol.~21, no.~3, pp. 2334--2360, 3rd Quart. 2019.

\bibitem{J_Morteza_Performance_2019}
M.~Banagar and H.~S. Dhillon, ``Performance characterization of canonical
  mobility models in drone cellular networks,'' \emph{IEEE Trans. Wireless
  Commun.}, vol.~19, no.~7, pp. 4994--5009, July 2020.

\bibitem{C_Fakhreddine_Handover_2019}
A.~Fakhreddine, C.~Bettstetter, S.~Hayat, R.~Muzaffar, and D.~Emini, ``Handover
  challenges for cellular-connected drones,'' in \emph{Proc. 5th Workshop on
  Micro Aerial Veh. Netw., Syst., Appl.}, June 2019, pp. 9--14.

\bibitem{C_Amer_Performance_2020}
R.~Amer, W.~Saad, B.~Galkin, and N.~Marchetti, ``Performance analysis of mobile
  cellular-connected drones under practical antenna configurations,'' in
  \emph{Proc. IEEE Int. Conf. Commun. (ICC)}, June 2020, pp. 1--7.

\bibitem{J_Nosratinia_Cooperative_2004}
A.~{Nosratinia}, T.~E. {Hunter}, and A.~{Hedayat}, ``Cooperative communication
  in wireless networks,'' \emph{IEEE Commun. Mag.}, vol.~42, no.~10, pp.
  74--80, Oct. 2004.

\bibitem{J_Laneman_Cooperative_2004}
J.~N. {Laneman}, D.~N.~C. {Tse}, and G.~W. {Wornell}, ``Cooperative diversity
  in wireless networks: Efficient protocols and outage behavior,'' \emph{IEEE
  Trans. Inf. Theory}, vol.~50, no.~12, pp. 3062--3080, Dec. 2004.

\bibitem{J_Hasna_Outage_2003}
M.~O. {Hasna} and M.~S. {Alouini}, ``Outage probability of multihop
  transmission over {Nakagami} fading channels,'' \emph{IEEE Commun. Lett.},
  vol.~7, no.~5, pp. 216--218, May 2003.

\bibitem{J_Hasna_End_2003}
------, ``End-to-end performance of transmission systems with relays over
  {Rayleigh}-fading channels,'' \emph{IEEE Trans. Wireless Commun.}, vol.~2,
  no.~6, pp. 1126--1131, Nov. 2003.

\bibitem{J_Cho_Throughput_2004}
J.~{Cho} and Z.~J. {Haas}, ``On the throughput enhancement of the downstream
  channel in cellular radio networks through multihop relaying,'' \emph{IEEE J.
  Sel. Areas Commun.}, vol.~22, no.~7, pp. 1206--1219, Sep. 2004.

\bibitem{J_Senaratne_Unified_2010}
D.~{Senaratne} and C.~{Tellambura}, ``Unified exact performance analysis of
  two-hop amplify-and-forward relaying in {Nakagami} fading,'' \emph{IEEE
  Trans. Veh. Technol.}, vol.~59, no.~3, pp. 1529--1534, Mar. 2010.

\bibitem{J_Aalo_Performance_2014}
V.~A. {Aalo} \emph{et~al.}, ``Performance analysis of multi-hop
  amplify-and-forward relaying systems in {Rayleigh} fading channels with a
  {Poisson} interference field,'' \emph{IEEE Trans. Wireless Commun.}, vol.~13,
  no.~1, pp. 24--35, Jan. 2014.

\bibitem{J_Lu_Stochastic_2015}
W.~{Lu} and M.~{Di Renzo}, ``Stochastic geometry modeling and system-level
  analysis \& optimization of relay-aided downlink cellular networks,''
  \emph{IEEE Trans. Commun.}, vol.~63, no.~11, pp. 4063--4085, Nov. 2015.

\bibitem{J_Zhang_Joint_2018}
S.~Zhang, H.~Zhang, Q.~He, K.~Bian, and L.~Song, ``Joint trajectory and power
  optimization for {UAV} relay networks,'' \emph{IEEE Commun. Lett.}, vol.~22,
  no.~1, pp. 161--164, Jan. 2018.

\bibitem{J_Chen_Multiple_2018}
Y.~Chen, N.~Zhao, Z.~Ding, and M.-S. Alouini, ``Multiple {UAVs} as relays:
  Multi-hop single link versus multiple dual-hop links,'' \emph{IEEE Trans.
  Wireless Commun.}, vol.~17, no.~9, pp. 6348--6359, Sep. 2018.

\bibitem{J_Chen_Optimum_2018}
Y.~Chen, W.~Feng, and G.~Zheng, ``Optimum placement of {UAV} as relays,''
  \emph{IEEE Commun. Lett.}, vol.~22, no.~2, pp. 248--251, Feb. 2018.

\bibitem{J_Pourranjbar_Novel_2020}
A.~{Pourranjbar}, M.~{Baniasadi}, A.~{Abbasfar}, and G.~{Kaddoum}, ``A novel
  distributed algorithm for phase synchronization in unmanned aerial
  vehicles,'' \emph{IEEE Commun. Lett.}, vol.~24, no.~10, pp. 2260--2264, Oct.
  2020.

\bibitem{B_Haenggi_Stochastic_2012}
M.~Haenggi, \emph{Stochastic Geometry for Wireless Networks}.\hskip 1em plus
  0.5em minus 0.4em\relax Cambridge, U.K.: Cambridge University Press, 2012.

\bibitem{J_Dhillon_Modeling_2012}
H.~S. Dhillon, R.~K. Ganti, F.~Baccelli, and J.~G. Andrews, ``Modeling and
  analysis of {K}-tier downlink heterogeneous cellular networks,'' \emph{IEEE
  J. Sel. Areas Commun.}, vol.~30, no.~3, pp. 550--560, Apr. 2012.

\bibitem{B_Dhillon_Poisson_2020}
H.~S. Dhillon and V.~V. Chetlur, \emph{Poisson Line {Cox} Process: Foundations
  and Applications to Vehicular Networks}.\hskip 1em plus 0.5em minus
  0.4em\relax Vermont, USA: Morgan \& Claypool Publishers, June 2020.

\bibitem{B_Morteza_Stochastic_2020}
M.~Banagar, V.~V. Chetlur, and H.~S. Dhillon, \emph{Stochastic Geometry-Based
  Performance Analysis of Drone Cellular Networks}.\hskip 1em plus 0.5em minus
  0.4em\relax John Wiley \& Sons, Ltd, 2020, ch.~9, pp. 231--254.

\bibitem{J_Morteza_Handover_2020}
------, ``Handover probability in drone cellular networks,'' \emph{IEEE
  Wireless Commun. Lett.}, vol.~9, no.~7, pp. 933--937, July 2020.

\bibitem{J_Amer_Mobility_2020}
R.~{Amer}, W.~{Saad}, and N.~{Marchetti}, ``Mobility in the sky: Performance
  and mobility analysis for cellular-connected {UAV}s,'' \emph{IEEE Trans.
  Commun.}, vol.~68, no.~5, pp. 3229--3246, May 2020.

\bibitem{C_Morteza_3GPP_2019}
M.~Banagar and H.~S. Dhillon, ``{3GPP}-inspired stochastic geometry-based
  mobility model for a drone cellular network,'' in \emph{Proc. IEEE Global
  Commun. Conf. (Globecom)}, Dec. 2019.

\bibitem{C_Morteza_Fundamentals_2019}
------, ``Fundamentals of drone cellular network analysis under random waypoint
  mobility model,'' in \emph{Proc. IEEE Global Commun. Conf. (Globecom)}, Dec.
  2019.

\bibitem{J_Morteza_Impact_2020}
M.~Banagar, H.~S. Dhillon, and A.~F. Molisch, ``Impact of {UAV} wobbling on the
  air-to-ground wireless channel,'' \emph{IEEE Trans. Veh. Technol.}, vol.~69,
  no.~11, pp. 14\,025--14\,030, Nov. 2020.

\bibitem{J_Chetlur_Downlink_2017}
V.~V. Chetlur and H.~S. Dhillon, ``Downlink coverage analysis for a finite
  3-{D} wireless network of unmanned aerial vehicles,'' \emph{IEEE Trans.
  Commun.}, vol.~65, no.~10, pp. 4543--4558, Oct. 2017.

\bibitem{J_Enayati_mobile_2018}
S.~{Enayati}, H.~{Saeedi}, H.~{Pishro-Nik}, and H.~{Yanikomeroglu}, ``Moving
  aerial base station networks: A stochastic geometry analysis and design
  perspective,'' \emph{IEEE Trans. Wireless Commun.}, vol.~18, no.~6, pp.
  2977--2988, June 2019.

\bibitem{J_Wang_Modeling_2018}
X.~{Wang} \emph{et~al.}, ``Modeling and analysis of aerial base
  station-assisted cellular networks in finite areas under {LoS} and {NLoS}
  propagation,'' \emph{IEEE Trans. Wireless Commun.}, vol.~17, no.~10, pp.
  6985--7000, Oct. 2018.

\bibitem{J_Alzenad_Coverage_2019}
M.~Alzenad and H.~Yanikomeroglu, ``Coverage and rate analysis for vertical
  heterogeneous networks ({VHetNets}),'' \emph{IEEE Trans. Wireless Commun.},
  vol.~18, no.~12, pp. 5643--5657, Dec. 2019.

\bibitem{J_Cherif_Downlink_2021}
N.~Cherif, M.~Alzenad, H.~Yanikomeroglu, and A.~Yongacoglu, ``Downlink coverage
  and rate analysis of an aerial user in vertical heterogeneous networks
  ({VHetNets}),'' \emph{IEEE Trans. Wireless Commun.}, vol.~20, no.~3, pp.
  1501--1516, Mar. 2021.

\bibitem{J_Azari_UAV_2020}
M.~M. Azari, G.~Geraci, A.~Garcia-Rodriguez, and S.~Pollin, ``{UAV-to-UAV}
  communications in cellular networks,'' \emph{IEEE Trans. Wireless Commun.},
  vol.~19, no.~9, pp. 6130--6144, Sep. 2020.

\bibitem{J_Galkin_Stochastic_2019}
B.~Galkin, J.~Kibilda, and L.~A. DaSilva, ``A stochastic model for {UAV}
  networks positioned above demand hotspots in urban environments,'' \emph{IEEE
  Trans. Veh. Technol.}, vol.~68, no.~7, pp. 6985--6996, Jul. 2019.

\bibitem{J_Dhillon_Backhaul_2015}
H.~S. {Dhillon} and G.~{Caire}, ``Wireless backhaul networks: Capacity bound,
  scalability analysis and design guidelines,'' \emph{IEEE Trans. Wireless
  Commun.}, vol.~14, no.~11, pp. 6043--6056, Nov. 2015.

\bibitem{J_Jaber_Wireless_2018}
M.~Jaber \emph{et~al.}, ``Wireless backhaul: Performance modeling and impact on
  user association for {5G},'' \emph{IEEE Trans. Wireless Commun.}, vol.~17,
  no.~5, pp. 3095--3110, May 2018.

\bibitem{J_Saha_Millimeter_2019}
C.~Saha and H.~S. Dhillon, ``Millimeter wave integrated access and backhaul in
  {5G}: Performance analysis and design insights,'' \emph{IEEE J. Sel. Areas
  Commun.}, vol.~37, no.~12, pp. 2669--2684, Dec. 2019.

\bibitem{J_Kishk_Aerial_2020}
M.~Kishk, A.~Bader, and M.-S. Alouini, ``Aerial base station deployment in {6G}
  cellular networks using tethered drones: The mobility and endurance
  tradeoff,'' \emph{IEEE Veh. Technol. Mag.}, vol.~15, no.~4, pp. 103--111,
  Dec. 2020.

\bibitem{C_Kouzayha_Stochastic_2020}
N.~Kouzayha \emph{et~al.}, ``Stochastic geometry analysis of hybrid aerial
  terrestrial networks with {mmWave} backhauling,'' in \emph{Proc. IEEE Int.
  Conf. Commun. (ICC)}, June 2020, pp. 1--7.

\bibitem{C_Galkin_Backhaul_2018}
B.~Galkin, J.~Kibilda, and L.~A. DaSilva, ``Backhaul for low-altitude {UAVs} in
  urban environments,'' in \emph{Proc. IEEE Int. Conf. Commun. (ICC)}, May
  2018, pp. 1--6.

\bibitem{J_Sabzehali_Optimizing_2021}
J.~Sabzehali \emph{et~al.}, ``Optimizing number, placement, and backhaul
  connectivity of multi-{UAV} networks,'' \emph{arXiv preprint}, Nov. 2021,
  [Online]. Available: https://arxiv.org/abs/2111.05457.

\bibitem{J_Chowdhury_3D_2020}
M.~M.~U. Chowdhury \emph{et~al.}, ``{3-D} trajectory optimization in
  {UAV}-assisted cellular networks considering antenna radiation pattern and
  backhaul constraint,'' \emph{IEEE Trans. Aerosp. Electron. Syst.}, vol.~56,
  no.~5, pp. 3735--3750, Oct. 2020.

\bibitem{J_Gapeyenko_Flexible_2018}
M.~Gapeyenko \emph{et~al.}, ``Flexible and reliable {UAV}-assisted backhaul
  operation in {5G mmWave} cellular networks,'' \emph{IEEE J. Sel. Areas
  Commun.}, vol.~36, no.~11, pp. 2486--2496, Nov. 2018.

\bibitem{J_Javad_3D_2021}
J.~Sabzehali, V.~K. Shah, H.~S. Dhillon, and J.~H. Reed, ``{3D} placement and
  orientation of {mmWave}-based {UAVs} for guaranteed {LoS} coverage,''
  \emph{IEEE Wireless Commun. Lett.}, to appear.

\bibitem{J_Amer_Toward_2019}
R.~Amer, W.~Saad, and N.~Marchetti, ``Toward a connected sky: Performance of
  beamforming with down-tilted antennas for ground and {UAV} user
  co-existence,'' \emph{IEEE Commun. Lett.}, vol.~23, no.~10, pp. 1840--1844,
  Oct. 2019.

\bibitem{3gpp_36873}
3GPP, ``{Study on 3D channel model for LTE},'' {3rd Generation Partnership
  Project (3GPP)}, Tech. Rep. 36.873, 12 2017.

\bibitem{3gpp_22829}
------, ``{Enhancement for unmanned aerial vehicles},'' {3rd Generation
  Partnership Project (3GPP)}, Tech. Rep. 22.829, 11 2018.

\bibitem{B_Stutzman_Antenna_2012}
W.~L. Stutzman and G.~A. Thiele, \emph{Antenna Theory and Design}.\hskip 1em
  plus 0.5em minus 0.4em\relax John Wiley \& Sons, Inc., May 2012.

\bibitem{J_Maeng_Base_2021}
S.~J. Maeng, M.~M.~U. Chowdhury, I.~Guvenc, A.~Bhuyan, and H.~Dai, ``Base
  station antenna uptilt optimization for cellular-connected drone corridors,''
  \emph{arXiv preprint}, July 2021, [Online]. Available:
  https://arxiv.org/abs/2107.00802.

\bibitem{J_Chowdhury_Ensuring_2021}
M.~M.~U. Chowdhury, I.~Guvenc, W.~Saad, and A.~Bhuyan, ``Ensuring reliable
  connectivity to cellular-connected {UAVs} with uptilted antennas and
  interference coordination,'' \emph{arXiv preprint}, Aug. 2021, [Online].
  Available: https://arxiv.org/abs/2108.05090.

\bibitem{3gpp_36777}
3GPP, ``{Study on enhanced LTE support for aerial vehicles},'' {3rd Generation
  Partnership Project (3GPP)}, Tech. Rep. 36.777, 01 2018.

\bibitem{J_AlHourani_Optimal_2014}
A.~Al-Hourani, S.~Kandeepan, and S.~Lardner, ``Optimal {LAP} altitude for
  maximum coverage,'' \emph{IEEE Wireless Commun. Lett.}, vol.~3, no.~6, pp.
  569--572, Dec. 2014.

\bibitem{B_Gould_Combinatorial_2010}
H.~W. Gould, \emph{Combinatorial Identities: Table I: Intermediate Techniques
  for Summing Finite Series}, May 2010, vol.~4.

\bibitem{J_Saha_3GPP_2018}
C.~{Saha}, M.~{Afshang}, and H.~S. {Dhillon}, ``{3GPP}-inspired {HetNet} model
  using {Poisson} cluster process: Sum-product functionals and downlink
  coverage,'' \emph{IEEE Trans. Commun.}, vol.~66, no.~5, pp. 2219--2234, May
  2018.

\bibitem{J_Rajanna_Downlink_2017}
A.~Rajanna and M.~Haenggi, ``Downlink coordinated joint transmission for mutual
  information accumulation,'' \emph{IEEE Wireless Commun. Lett.}, vol.~6,
  no.~2, pp. 198--201, Apr. 2017.

\end{thebibliography}
\end{document}